\documentclass[aps,twocolumn,floats,superscriptaddress,prd,nofootinbib]{revtex4-1}
\usepackage{graphicx, epsfig, bm, amsmath}

\usepackage{color}
\usepackage{hyperref}
\usepackage{ wasysym }
\begin{document}


\newcommand{\lcdm}{$\Lambda$CDM}

\newcommand{\gpr}{G^{\prime}}

\newcommand{\fnl}{f_{\rm NL}}
\newcommand{\curv}{{\cal R}}

\definecolor{darkgreen}{cmyk}{0.85,0.2,1.00,0.2}
\newcommand{\peter}[1]{\textcolor{red}{[{\bf PA}: #1]}}
\newcommand{\cora}[1]{\textcolor{darkgreen}{[{\bf CD}: #1]}}
\newcommand{\wh}[1]{\textcolor{blue}{[{\bf WH}: #1]}}
\newcommand{\gB}{g_B}
\newcommand{\WP}{W}
\newcommand{\XP}{X}
\newcommand{\B}{B^{\rm Bulk}}

\newcommand{\aap}{Astron. Astrophys.}


\pagestyle{plain}

\title{Fast Computation of Bispectrum Features with Generalized Slow Roll}

\author{Peter Adshead}
\affiliation{Kavli Institute for Cosmological Physics,  Enrico Fermi Institute, University of Chicago, Chicago, Illinois 60637, USA}
        
\author{  Wayne Hu}
\affiliation{Kavli Institute for Cosmological Physics,  Enrico Fermi Institute, University of Chicago, Chicago, Illinois 60637, USA}
\affiliation{Department of Astronomy \& Astrophysics, University of Chicago, Chicago, Illinois 60637, USA}
        
\author{Cora Dvorkin}
\affiliation{Kavli Institute for Cosmological Physics,  Enrico Fermi Institute, University of Chicago, Chicago, Illinois 60637, USA}
\affiliation{Department of Physics, University of Chicago, Chicago Illinois 60637, USA}

\author{Hiranya V. Peiris}
\affiliation{Department of Physics and Astronomy, University College London, London WC1E 6BT, U.K.}
\affiliation{Institute of Astronomy and Kavli Institute for Cosmology, University of Cambridge, Cambridge CB3 0HA, U.K.}

\begin{abstract}
We develop a fast technique  based on the generalized slow-roll (GSR) approach for computing the curvature bispectrum of inflationary models with features.  We show that all triangle configurations can be expressed in terms of three simple integrals over the inflationary background with typical accuracy of better than $\sim 20\%$. 
With a first order GSR approach the typical accuracy can be improved to better than the $\sim5\%$ level.    We illustrate this technique with the step potential model that has
been invoked to explain the WMAP temperature power spectrum glitches at $\ell \sim 20-40$ and show that the maximum likelihood model falls short of observability by more than a factor of 100 in amplitude. We also explicitly demonstrate that the bispectrum consistency relation with the local slope of the power spectrum is satisfied for these models.
 In the GSR approach, the bispectrum arises from integrals of nearly the same function of the background slow-roll parameters as the power spectrum but with a stronger weight to the epoch before horizon crossing.  Hence this technique enables reverse engineering of models with large bispectrum but small power spectrum features.
\end{abstract}

\maketitle

\section{Introduction}
\label{sec:intro}

In this paper, we develop the generalized slow-roll (GSR) approach to obtain the bispectrum of curvature fluctuations produced by features in the inflaton potential. Bispectra for these kinds of models have been previously considered
by a computationally intensive direct integration of the curvature fluctuations for each configuration \cite{Chen:2006xjb,Chen:2008wn}.   The GSR approach provides a computationally efficient method that involves only a single function of the inflationary background.

The GSR approximation was originally introduced  to allow for accurate solutions of the power spectrum for models where the slow-roll parameters are small but not necessarily constant \cite{Stewart:2001cd}. This method was 
subsequently extended  for cases in which the potential can have large features and the slow-roll parameters are not small \cite{Dvorkin:2009ne}.

Violations of slow roll arise when the inflaton traverses a feature, such as a step or a bump, in its potential. As the inflaton rolls across the feature, some of its potential energy is converted into kinetic energy or vice versa. By keeping the amplitude of the feature small, one is able to ensure that inflation is not interrupted; however, by arranging for the feature to be sharp, the inflaton undergoes a sharp transient acceleration which temporarily violates slow roll.  

The possibility of sharp features in the inflationary potential has a long history. Starobinsky first discussed the spectrum of adiabatic fluctuations for a potential with sharp features \cite{Starobinsky:1992ts}.
In particular, sudden downward step features can arise naturally in models of inflation derived from supergravity \cite{Adams:1997de}.  Detailed numerical analysis showed that these features in the potential lead to oscillating features in the spectrum of curvature fluctuations \cite{Adams:2001vc} and consequently, such models have been invoked to explain glitches observed in the temperature power spectrum at scales around $\ell \sim 20 -40$ \cite{Peiris:2003ff,Covi:2006ci,Hamann:2007pa, Mortonson:2009qv}. 
Although  the bispectrum for such models has been previously studied \cite{Chen:2006xjb,Chen:2008wn}, the intensive computation required has prevented a full assessment of its observability. The authors in  \cite{Chen:2006xjb,Chen:2008wn} wrote down an approximate analytic form for the bispectrum produced by a step but did not pursue it further.  Consequently we use this step model to illustrate the GSR bispectrum technique. While we find that  the bispectrum produced by a step with the parameters chosen to best fit the glitch at $\ell  =20-40$ is unobservable, by abandoning this prior an adjusting the width and height of the step, larger bispectra can be produced \cite{Chen:2006xjb}. 

Features have also been shown to arise from other phenomenological processes:
duality cascades \cite{Hailu:2006uj,Bean:2008na}, waterfall transitions \cite{Abolhasani:2010kn}, the imprints of heavy physics on the inflaton \cite{Achucarro:2010da}, fast phase transitions \cite{Joy:2007na}, and multiple field scenarios \cite{Hotchkiss:2009pj}.  Sudden changes in the sound velocity in more general inflationary models have also been shown to give rise to such features \cite{Nakashima:2010sa}. The techniques we develop here apply only to cases where inflation is being driven by a single effective degree of freedom. The crucial assumption is that there is only one ``clock." We work with canonical kinetic terms but expect the generalization to other forms to be straightforward.

The outline of this paper is as follows.  In \S \ref{sec:methods} we review
the formalism for computing the bispectrum in the literature.  We derive
the GSR approximation for the bispectrum in Appendix \ref{app:GSR} and
apply it to the step model in \S \ref{sec:GSRbi}.  In Appendix \ref{app:eugene}, we test our calculations against results in the literature.   We use the fast
GSR approximation in \S \ref{sec:applications} to estimate the observability of
the bispectrum step features for a model that fits the WMAP power spectrum
glitches.  
Throughout we work in units where the reduced Planck mass $M_{\rm Pl} =(8\pi G)^{-1/2}= 1$.

\section{Bispectrum Formalism}
\label{sec:methods}
 In this section we begin by briefly outlining the general method used to evaluate the bispectrum employed by Maldacena \cite{Maldacena:2002vr} and extended by Weinberg \cite{Weinberg:2005vy}.
   

We work in comoving gauge, where the time slicing is chosen so that the scalar fluctuations are in the metric, and make use of the interaction picture, where these curvature fluctuations $\mathcal{R}$ evolve according to the equations of motion derived from the quadratic action
\begin{align}\label{eqn:quadaction}
S_{2} = \frac{1}{2}\int dt d^{3}x\,a^{3}2\epsilon_{H}\left[\mathcal{\dot{R}}^{2} -\frac{ (\partial\mathcal{R})^2}{a^{2}}\right].
\end{align}
In this picture, the three-point function arises from interaction terms defined to leading order by the cubic action 
 \cite{Maldacena:2002vr}
\begin{align}\label{eqn:action3rd}\nonumber
S_{3} =&  \int dt d^{3}x \bigg[ a^{3}\epsilon_{H}^2 \mathcal{R}\dot{\mathcal{R}}^2 + a \epsilon_{H}^{2}\mathcal{R}(\partial \mathcal{R})^{2} - 2 a \epsilon_{H} \dot{\mathcal{R}}\partial\mathcal{R}\partial\chi \\\nonumber & + a^{3}\epsilon_{H}(\dot \epsilon_{H}-\dot \eta_{H})\mathcal{R}^{2}\dot{\mathcal{R}} + \frac{\epsilon_{H}}{2 a}\partial\mathcal{R}\partial\chi\partial^{2}\chi\\ &-\frac{d}{dt}\left(a^{3}2\epsilon_{H} \mathcal{\dot{R}} f(\mathcal{R})\right)
\bigg],
\end{align}
which evolve the states.
Here
\begin{align}\nonumber
\eta_{H} = & -\frac{\ddot{\phi}}{\dot{\phi}H},\quad \epsilon_{H} = \frac{\dot{\phi}^{2}}{2H^{2}},\quad \chi =  a^{2}\epsilon_{H}\partial^{-2}\dot{\mathcal{R}},
\\
&f(\mathcal{R}) =- \frac{1}{2}(\epsilon_{H}-\eta_{H})\mathcal{R}^2+\ldots,
\end{align}
overdots are derivatives with respect to cosmic time $t$, $\partial_{i}$ refers to the spatial derivative, spatial indices here are contracted with the Kronecker delta, $\partial^{2} = \delta^{ij}\partial_{i}\partial_{j}$ , and `$\ldots$' denotes terms which vanish outside the horizon.

The form of the action in Eq.~(\ref{eqn:action3rd}) is slightly different from that originally written down by Maldacena \cite{Maldacena:2002vr}. Maldacena omits the total derivative terms, while including terms proportional to the first order equations of motion which are of the form $f(\curv)\delta \mathcal{L}_{2}/\delta \mathcal{R}$. However, since these terms are evaluated on-shell, their contribution to any Feynman graph is formally zero. Maldacena removes these terms by performing a field redefinition to a new variable $\curv_{n} = \curv - f(\curv)$. As described in \cite{AdsheadSeery}, the inclusion of the boundary term in the last line of Eq.~(\ref{eqn:action3rd}) accounts for the terms one obtains by performing the field redefinition.

We quantize the theory in the usual way.
The translational invariance of the background makes it convenient to expand the field $\mathcal{R}$ in Fourier components, 
\begin{align}
\mathcal{R}({\bf x}, t) = \int \frac{d^{3}q}{(2\pi)^3}e^{i {\bf q}\cdot{\bf x}}\mathcal{R}_{{\bf q}}( t).
\end{align}
Rotational invariance and Hermiticity imply that the most general solution takes the form
\begin{align}
\hat{\mathcal{R}}_{{\bf q}}( t) = \mathcal{R}_{ q}( t)\hat a({\bf q}) + \mathcal{R}^{*}_{ q}( t)\hat a^{\dagger}(-{\bf q}),
\end{align}
where $q = |{\bf q}|$ is the magnitude of the comoving momentum and hats 
denote operators. In the quantum theory, $\hat{a}({\bf q})$ and $\hat{a}^{\dagger}({\bf q})$ can be thought of as creation and annihilation operators satisfying
\begin{align}
[\hat a({\bf q}), \hat a^{\dagger}({\bf q}')] = (2\pi)^{3}\delta({\bf q} - {\bf q}').
\end{align}
With these definitions, we can construct the Fock space by applying creation operators to the state annihilated by all of the $\hat{a}({\bf q})$ which we call the vacuum, $|0\rangle$. We can then define the two-point function, or propagator,
\begin{align}\label{eqn:propagator}
\langle \hat{ \mathcal{R}}_{\bf q}(t)\hat{\mathcal{R}}_{\bf q'}(t')\rangle = \mathcal{R}_{q}(t)\mathcal{R}^{*}_{q'}(t')(2\pi)^{3}\delta({\bf q}+{\bf q}'),
\end{align}
which follows by simply normal ordering. The prescription of the ``in-in'' formalism for the expectation value of a product of field operators $O(t)$ is to evaluate the expression
\begin{align}\label{eqn:inin}
\langle O(t)\rangle = \langle U^{\dagger}(t, t_{0})O(t)U(t, t_{0})\rangle,
\end{align}
where $U(t, t_{0})$ is the time evolution operator
\begin{align}
U(t, t_{0}) = T\exp\left( - i\int_{t_{0}}^{t}H_{I}(t) dt\right).
\end{align}
The average in Eq.~(\ref{eqn:inin}) denoted by $\langle ... \rangle$ is a quantum average with respect to the vacuum state of the free field theory governed by the action in Eq.~(\ref{eqn:quadaction}).
For the problem at hand, we take the initial time $t_{0}$ to be in the asymptotic past, $t_{0} = -\infty(1+i\varepsilon)$, where the $i\varepsilon$ prescription projects out the Bunch-Davies state initially.

The tree-level bispectrum is then given by expanding Eq.~(\ref{eqn:inin}) [with $O(t) = \mathcal{R}_{\bf k_{1}}(t_{*})\mathcal{R}_{\bf k_{2}}(t_{*})\mathcal{R}_{\bf k_{3}}(t_{*})$] to linear order
\begin{align}\label{eqn:bispec}\nonumber
& \langle\hat{\mathcal{R}}_{\bf k_{1}}(t_{*})\hat{\mathcal{R}}_{\bf k_{2}}(t_{*})\hat{\mathcal{R}}_{\bf k_{3}}(t_{*})\rangle = \\& 2\Re\left[ -i\int^{t_{*}}_{-\infty} dt \langle \hat{\mathcal{R}}_{\bf k_{1}}(t_{*})\hat{\mathcal{R}}_{\bf k_{2}}(t_{*})\hat{\mathcal{R}}_{\bf k_{3}}(t_{*})H_{I}(t)\rangle\right].
\end{align}

In this work we are interested in potentials in which the inflaton undergoes a sharp transient acceleration but inflation is not interrupted. Consequently, $\epsilon_{H}\ll1$ everywhere and  the bispectrum is dominated by the term proportional to the  derivative of $\eta_{H}$. To a very good approximation, for the purposes of this paper, the cubic action defined in Eq. (\ref{eqn:action3rd}) reduces to
\begin{align}\nonumber\label{eqn:truncatedaction}
S_{3} \approx & \int dt d^{3}x \, \bigg[a^{3}\epsilon_H (\dot\epsilon_H - \dot\eta_H)\mathcal{R}^{2}\dot{\mathcal{R}} \nonumber\\
&\qquad 
- \frac{d}{dt}\left(a^{3}\epsilon_H(\epsilon_H-\eta_H)\mathcal{R}^{2}\dot{\mathcal{R}}\right)\bigg].
\end{align}
For this cubic action in Eq.~(\ref{eqn:action3rd}), the interaction Hamiltonian is
\begin{align}\nonumber\label{eqn:Hamiltonian}
H_{I}(t) =&  -\int d^{3}x  \Bigg[a^{3}\epsilon_H (\dot \epsilon_H-\dot\eta_H)\hat{\mathcal{R}}^{2}\dot{\hat{\mathcal{R}}}\\ 
&-\frac{d}{dt}\left(a^{3}\epsilon_H(\epsilon_H-\eta_H)\hat{\mathcal{R}}^{2}\dot{\hat{\mathcal{R}}}\right)\Bigg].
\end{align}
Switching to conformal time, $\eta = \int_{t}^{t_{\rm end}}dt'/a$ (defined to be a positive quantity during inflation) and working in Fourier space, we obtain for the interaction Hamiltonian 
\begin{align}
H_{I}(\eta) = - \int&\frac{d^{3}q_a}{(2\pi)^3}\frac{d^{3}q_b}{(2\pi)^3}\frac{d^{3}q_c}{(2\pi)^3} (2\pi)^{3}\delta^{3}({\bf q}_a+{\bf q}_b+{\bf q}_c)\nonumber\\ 
& \times\Bigg[ \frac{a^{2}\epsilon_{H}}{3\eta^2}   (\epsilon_{H} - \eta_{H})' \left(  \hat{\mathcal{R}}_{{\bf q}_a}\hat{\mathcal{R}}_{{\bf q}_b}\hat{\mathcal{R}}_{{\bf q}_c}\right)' \\
\nonumber
& -\frac{d}{d\eta}\left(\frac{a^{2} \epsilon_{H}}{3\eta}(\epsilon_{H}-\eta_{H})(\hat{\mathcal{R}}_{{\bf q}_a}\hat{\mathcal{R}}_{{\bf q}_b}\hat{\mathcal{R}}_{{\bf q}_c})'\right)\Bigg],
\end{align}
where here and throughout $' \equiv d/d\ln\eta$.
In this expression, the fields $\mathcal{R}$ are interaction picture fields whose time dependence is governed by the Hamiltonian derived from the quadratic action in Eq.~(\ref{eqn:quadaction}). 

We are only interested in the connected part of the three-point function here, since this is the only part that contributes to the non-Gaussianity. This is evaluated from Eq.~(\ref{eqn:bispec}) using Eq.~(\ref{eqn:propagator}) and Wick's theorem. 
Defining the bispectrum through
\begin{equation}
\langle \hat{\curv}_{{\bf k}_1} \hat{\curv}_{{\bf k}_2} \hat{\curv}_{{\bf k}_3}\rangle
= (2\pi)^3 \delta({\bf k}_1+ {\bf k}_2+{\bf k}_3) B_{\curv}(k_1,k_2,k_3),
\end{equation}
we find
\begin{align}\nonumber\label{eqn:bispectrum1st}
& B_{\curv}(k_1,k_2,k_3) =  4 \Re\Bigg\{ i\mathcal{R}_{k_{1}}(\eta_{*})\mathcal{R}_{k_{2}}(\eta_{*})\mathcal{R}_{k_{3}}(\eta_{*})\\ \nonumber
&\times \Bigg[ \int_{\eta_{*}}^{\infty} {d\eta \over \eta^2}\,  {a^{2} \epsilon_{H}}(\epsilon_{H} - \eta_{H})' (\mathcal{R}^{*}_{k_{1}}\mathcal{R}^{*}_{ k_{2}}\mathcal{R}^{*}_{ k_{3}})'\\
& +{a^{2}\epsilon_{H} \over \eta_*}(\epsilon_{H}-\eta_{H})(\mathcal{R}^{*}_{k_{1}}\mathcal{R}^{*}_{ k_{2}}\mathcal{R}^{*}_{k_{3}})' \Big|_{\eta = \eta_{*}} \Bigg]
\Bigg\}.
\end{align}

The bispectrum in Eq.~(\ref{eqn:bispectrum1st}) may appear to depend on the final time $\eta_{*}$.   However, if we consider modes that are well outside the horizon, integrate Eq.~(\ref{eqn:bispectrum1st})  by parts and use the equation of motion for $\mathcal{R}$,  we obtain
\begin{align}\label{eqn:bispectrum1stintbyparts}
& B_{\curv}(k_1,k_2,k_3) =  4 \Re\Bigg\{ i(k_{1}^2+k_{2}^2+k_{3}^2)\mathcal{R}_{k_{1}}\mathcal{R}_{k_{2}}\mathcal{R}_{k_{3}}\\ \nonumber
&\times \int_{\eta_{*}}^{\infty} {d\eta}\, {a^{2} \epsilon_{H}} (\epsilon_{H} - \eta_{H}) \mathcal{R}^{*}_{k_{1}}\mathcal{R}^{*}_{ k_{2}}\mathcal{R}^{*}_{ k_{3}}
\Bigg\}.
\end{align}
We have also dropped the contributions from terms like
\begin{align}
a^2\, \left({d\ln\curv_k\over d\ln\eta}\right)^2 = \mathcal{O}(k^4\eta^{4})
\end{align}
since these converge essentially as inverse powers of the scale factor for these modes. Since the modes under consideration are well outside the horizon, they have become constant, and we can pull them out of the integral and write this expression as
\begin{align}\label{eqn:superHcancellation}\nonumber
 B_{\curv}(k_1,k_2,k_3) =&  4 \Re\Bigg\{ i(k_{1}^2+k_{2}^2+k_{3}^2)|\mathcal{R}_{k_{1}}|^2|\mathcal{R}_{k_{2}}|^2|\mathcal{R}_{k_{3}}|^2\\ 
&\quad\times \int_{\eta_{*}}^{\infty} {d\eta}  \,{a^{2} \epsilon_{H}}(\epsilon_{H} - \eta_{H}) 
\Bigg\},
\end{align}
which is obviously identically zero, regardless of the time at which we choose its evaluation. The badly divergent integral cannot acquire an imaginary part. 

One might worry that this statement holds only at leading order. In writing down Eq.~(\ref{eqn:superHcancellation}) we have neglected the decaying mode which may be amplified by the divergent integrand if it were not decaying fast enough at late times. It is easy to see that this does not occur by examining the asymptotic expansion of the mode functions $\curv$ about the superhorizon limit \cite{Weinberg:2005vy},
\begin{align}\nonumber
\curv_k(\eta) = & \curv^{0}_{k}\left(1+k^{2}\int_{\eta}^{0} {d\eta' \over  a^2\epsilon_H}\int_{\infty}^{\eta'} d\eta''\,a^2\epsilon_{H}\right)\\\nonumber
&+\mathcal{A}_{k}\int_{\eta}^{0}{d\eta'\over a^{2}\epsilon_{H}}\left(1+k^{2}\!\int_{\eta'}^{0} \!\!{d\eta'' \over a^2 \epsilon_H }\int_{\infty}^{\eta''} d\tilde\eta\,a^2\epsilon_{H}\!\right)\!\!\\ & +\ldots,
\end{align}
where $ \curv^{0}_{k}$ and $\mathcal{A}_{k}$ are complex constants and `$\ldots$' refers to terms higher order in $k$ in the expansion. Thus, the leading order corrections to Eq.~(\ref{eqn:superHcancellation}) are proportional to
\begin{align}
\Im[\curv^{0}_{k}\mathcal{A}_{k}]\int_{\eta_{*}}^{\infty}d\eta\,{a^{2} \epsilon_{H}}(\epsilon_{H} - \eta_{H}) \int_{\eta}^{0}{d\eta'\over a^{2}\epsilon_{H}},
\end{align}
which, as long as the slow-roll parameters are well behaved, converges essentially as an inverse power of the scale factor. 

Upon examination of Eq.~(\ref{eqn:bispectrum1stintbyparts}), it can be seen that this conclusion is in fact a direct consequence of the general theorem proved by Weinberg \cite{Weinberg:2005vy}. Eq.~(\ref{eqn:bispectrum1stintbyparts}) can be thought of as arising from a ``dangerous interaction.'' That is, an interaction that diverges as $a$ at late times, but contains only fields and not time derivatives of fields. Then, due to Weinberg's theorem the integrals over the time coordinates of the interaction converge exponentially fast in cosmic time at late times. We also point out that the second term in Eq.~(\ref{eqn:action3rd}) takes this form, and we are safe in concluding that it does not give large or divergent contributions to correlation functions at late times.

This property shows that the time independence of the three-point function after horizon crossing is enforced by the boundary term regardless of how the slow-roll parameters are evolving.

 \section{Generalized Slow Roll}\label{sec:GSRbi}

The generalized slow-roll (GSR) approach introduced by Stewart \cite{Stewart:2001cd} is a technique
for predicting the curvature power spectrum in models where the inflation 
continues uninterrupted but the slow-roll parameters $\epsilon_H$ and $\eta_H$
evolve rapidly due to the presence of sharp features in the inflaton potential.

Here we extend the GSR approach in two ways.   First, we generalize the techniques
for the calculation of the bispectrum.  Second, we make them appropriate for the
calculation of large amplitude features as was done for the power spectrum in Ref.~\cite{Dvorkin:2009ne}.  The latter involves adding certain formally higher order terms, which
can be justified from
an iterative expansion as discussed in Appendix \ref{app:GSR}.  
We use this approach to develop a fast technique that can be used to approximate all 
configurations of the bispectrum for these kinds of models.

We present only the main results in this section, leaving the details of the derivation and an overview of the GSR technique to Appendix \ref{app:GSR}. In the subsections we present comparisons of our approximation to the exact computation in the equilateral, squeezed and flat limits.

\subsection{Zeroth-Order Expressions}
\label{sec:GSRzero}

The GSR approach proceeds by iteratively correcting the evolution of the mode function for the effect of deviations from de Sitter space
[see Eq.~(\ref{eqn:formal})].

At zeroth order, we employ only the de Sitter forms for the $\mathcal{R}_k$ mode functions in a specific way described in Appendix \ref{app:GSR}.  
 Eq.~(\ref{eqn:bispectrum1st}) then simplifies  to \begin{eqnarray}
B_{\curv}(k_1,k_2,k_3)  &\approx & {(2 \pi)^4  \over k_1^3 k_2^3 k_3^3} {\Delta_{\cal R}(k_1) 
\Delta_{\cal R}(k_2)
\Delta_{\cal R}(k_3)\over 4} \nonumber\\&&\quad
\Big[ -I_0(K) k_1 k_2 k_3 - I_1(K) \sum_{i \ne j} k_i^2 k_j 
\nonumber\\&&\quad + I_2(K) K (k_1^2+k_2^2 + k_3^2) \Big]
\label{eqn:gsrlbi}
\end{eqnarray}
involving integrals separable in
$k$, 
\begin{eqnarray}\label{eqn:Integrals}
I_0(K) &=&\int_{0}^{\infty} {d\eta \over \eta}  G_B'(\ln\eta) (K\eta) \sin (K\eta),\nonumber \\
I_1(K) &=&G_B(\ln\eta_*) + \int_{\eta_*}^{\infty}{ d\eta \over \eta}
 G_B'(\ln\eta) {\cos (K\eta)}, \\
I_2(K) &=&G_B(\ln\eta_*) +  \int_{\eta_*}^{\infty} { d\eta \over \eta} G_B'(\ln\eta) {\sin (K\eta) \over K\eta} ,\nonumber
\end{eqnarray}
which depend only on the perimeter of the triangle  $K=k_1+k_2+k_3$.  Thus the bispectra for all possible triangles can be efficiently obtained by precomputing these three integrals. Note that the trigonometric functions for $I_1$ and $I_2$ for $K\eta \ll 1$ approach unity and so the expressions become independent of the arbitrarily chosen end point $\eta_*$.  
While this cancellation is guaranteed by the form of the action in Eq.~(\ref{eqn:bispectrum1st}), it is not guaranteed to occur order by order in the GSR expansion of the mode functions.  This is due to the fact that in Eq.~(\ref{eqn:bispectrum1st}), the effects of the source are compensated by the response of the derivative of the curvature on super horizon scales. At zeroth order in the GSR approximation, the mode function does not respond to the feature at all, leaving this aspect of the source uncompensated. As we explain in Appendix \ref{app:GSR} we have included the appropriate higher order terms in the source function in order to enforce this cancellation.

The inclusion of higher order terms modifies the source from Eq.~(\ref{eqn:bispectrum1st})  to
\begin{equation}
G_B' =   \left( {\epsilon_H - \eta_H \over f} \right)',\quad {\rm with}\quad G_B=   \left( {\epsilon_H - \eta_H \over f} \right),
\end{equation} 
where  
\begin{equation}
f ={ \sqrt{ 8\pi^2 \epsilon_H} \over H}  (a H \eta) ,
\label{eqn:fdef}
\end{equation}
and $\Delta_{\cal R}^2 =  k^3 P_{\cal R} /2\pi^2$ is the curvature power spectrum.
Note that for constant $\epsilon_H \ll 1$, $aH \eta =1$ and $f^{-2} = {\Delta_{\cal R}^2}$ is the usual slow
roll result for the power spectrum.

The modification to the source and the replacement of the power spectrum 
for the zeroth-order external modes evaluated at $\eta_*$ in Eq.~(\ref{eqn:bispectrum1st}) represent the two higher order corrections we have introduced.  Both have the effect of enforcing that the bispectrum is insensitive to features in the inflaton potential for modes that are superhorizon scale when 
the inflaton crosses the feature.   For the source, this is achieved by making
it a exact derivative of a combination of slow-roll parameters and affects
mainly the $I_1$ and $I_2$ terms.

In other words, we enforce that the three-point correlation function of curvature perturbations remains  exactly constant outside the horizon at zeroth order in the GSR expansion of the mode functions. In Appendix \ref{app:GSR}, we show that the higher order terms we add to the zeroth-order approach are compatible with a fully first order GSR computation.

It is interesting to note the similarities and differences between how evolution in the slow-roll parameters affect the power spectrum and bispectrum.  The 
source to the power spectrum is [see Eq.~(\ref{eqn:GSRLfullpower})]
\begin{equation}
G' = {2 \over 3} \left( {f'' \over f} -3 {f' \over f} - {f'^2 \over f^2} \right) \approx
-{2\over 3} f G_B',
\end{equation}
where the approximation follows for cases where  $f''/f$ is the dominant 
contribution to both, as in the step potential below.  As noted above,  in the ordinary slow-roll 
approximation $f^{-2}$ is the amplitude of the power spectrum.  
Thus, the presence of the extra factor of $f$ serves to make the bispectrum scale
as the square of the power spectrum, leaving an otherwise similar source
for the power spectrum and bispectrum.

The main difference between the impact of features on the power spectrum and bispectrum is in the $I_0$ integral in Eq.~(\ref{eqn:Integrals}).   Unlike those for the
power spectrum, this integral carries a divergent $K\eta \sin(K\eta)$ 
term as $K\eta \rightarrow \infty$ which can be traced back to the appearance of derivatives of the
mode functions in Eq.~(\ref{eqn:bispectrum1st}). 

The implication is that deviations generate bispectrum correlations
while the modes are deeper within the horizon compared with the power spectrum.
Thus one generically expects that the impact of features in the inflaton 
potential on the bispectrum extends to higher $k$ than in the power spectrum,
enhancing their observability. 
\begin{figure}[t]
\centerline{\psfig{file=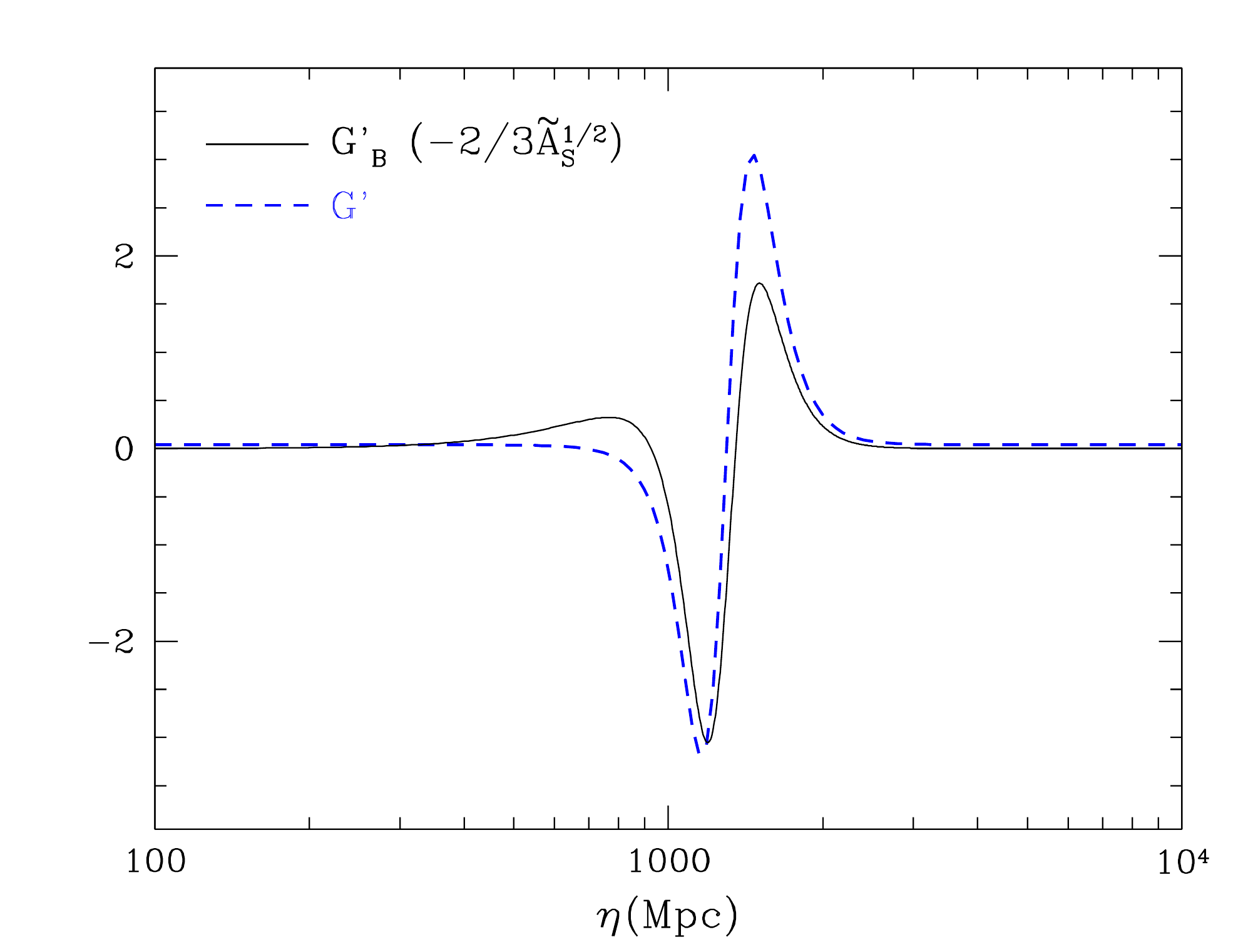, width=3.2in}}
\caption{\footnotesize Bispectrum source $G_B'$ and power spectrum source $G'$ in the GSR approximation for the step potential model (see \S \ref{sec:step}).  The bispectrum
source has been rescaled by a constant related to $\tilde A_S= 2.39\times 10^{-9}$ to eliminate its scaling with the power 
spectrum squared.  Both sources share similar structure with features that
integrate to zero as required for constant superhorizon behavior.}
\label{fig:source}
\end{figure}

\subsection{Step Potential}
\label{sec:step}

As an example, we consider the potential \cite{Adams:2001vc}
\begin{align}
V(\phi) = \frac{1}{2}m^{2}\phi^{2}\left[1+c\tanh\left(\frac{\phi - \phi_{s}}{d}\right)\right],
\end{align} 
which corresponds to a smooth step at $\phi = \phi_{s}$ of fractional height $c$ and width $d$. Such a feature in the inflationary potential has been invoked to  explain the `glitches' in the CMB temperature anisotropy data at $\ell = 20-40$ \cite{Covi:2006ci,Hamann:2007pa}.
Addressing the observability of the corresponding features
in the bispectrum requires a fast approach to their calculation
(see \S \ref{sec:applications}).

For concreteness we adopt the WMAP5 
maximum likelihood values of the parameters of the step potential \cite{Mortonson:2009qv} $\{m,c,d,\phi_{s}\} = \{7.126\times10^{-6}, 1.505\times 10^{-3}, 0.02705, 14.668 \}$. 
We plot the bispectrum and power spectrum source functions
for this model in Fig.~\ref{fig:source}.  Note their similar structure once rescaled
in amplitude.
In Appendix \ref{app:eugene} we consider an alternate choice of parameters to make contact with results in the 
literature.   

\begin{figure}[t]
\centerline{\psfig{file=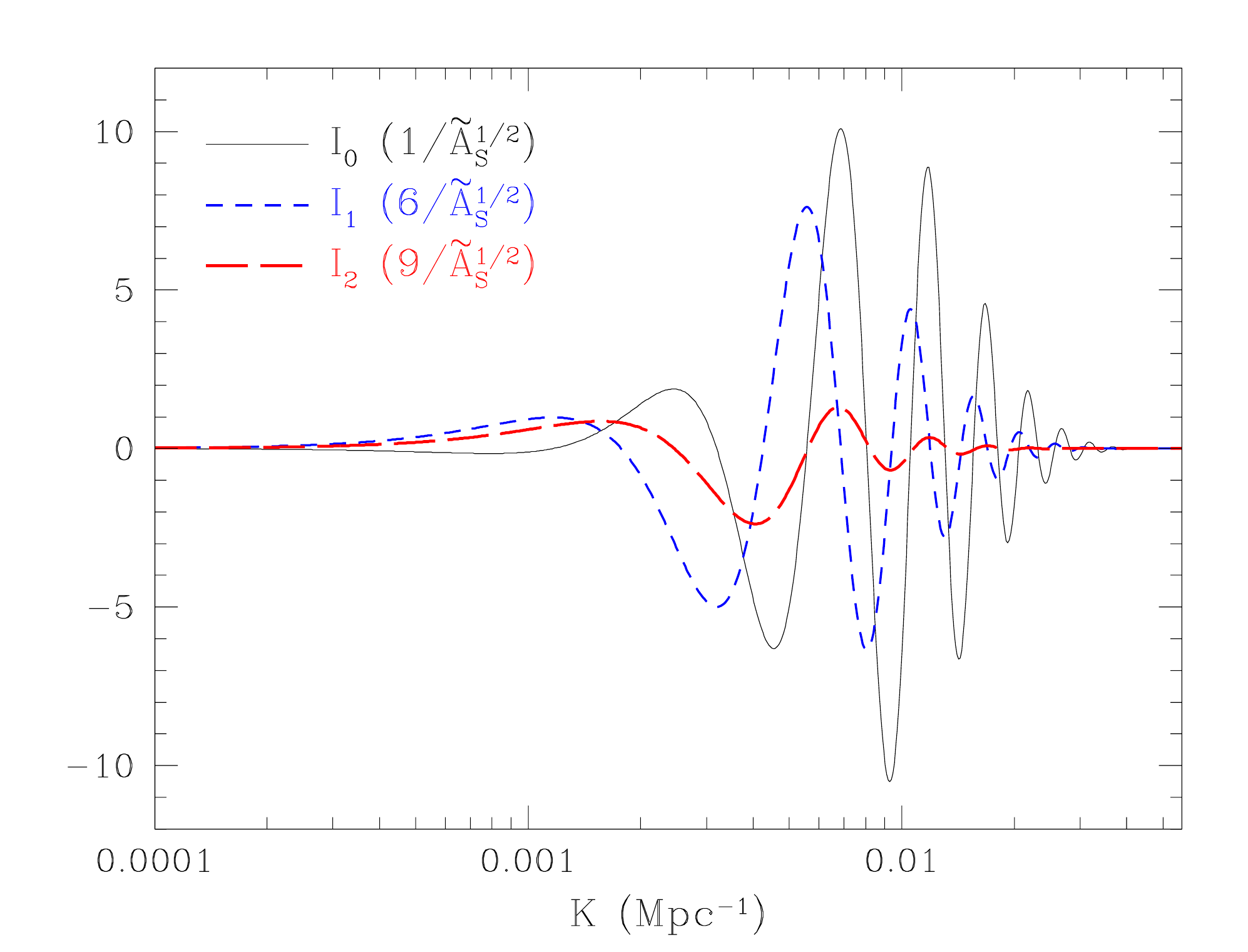, width=3.2in}}
\caption{\footnotesize Bispectrum integrals in the zeroth-order GSR approximation for the step
potential model as a function of the perimeter of the triangles $K=k_1+k_2+k_3$.
All bispectrum triangles can be formed efficiently from these three integrals.  Weights reflect how the integrals contribute to equilateral configurations. Note that $I_0$
dominates the high $k$ structure.}
\label{fig:integrals}
\end{figure}

Following the notation of the existing literature \cite{Chen:2006xjb,Chen:2008wn}, we construct plots related to 
\begin{align}
\mathcal{G}(k_{1}, k_{2},k_{3})= \frac{k_{1}^{3}k_{2}^{3}k_{3}^{3}}{(2\pi)^{4}\tilde{A}_{S}^2}B_{\curv}(k_1,k_2,k_3),
\label{eqn:dimensionlessbi}
\end{align}
where $\tilde{A}_{S}$ is an arbitrary constant that is of order the curvature
power spectrum normalization without the step feature.   
 In practice we take $\tilde A_S = 2.39\times 10^{-9}$. Since $\mathcal{G}$ has dimensions of $k^3$,  we typically divide it by some representative
$k^3$.
In  Fig.~\ref{fig:integrals} we show the one-dimensional GSR bispectrum integrals of Eq.~(\ref{eqn:Integrals}) for this model. Note that $I_{0}(K)$ 
dominates, especially at high $k$ as discussed above.

\subsection{Equilateral Limit}

\begin{figure}[t]
\centerline{\psfig{file=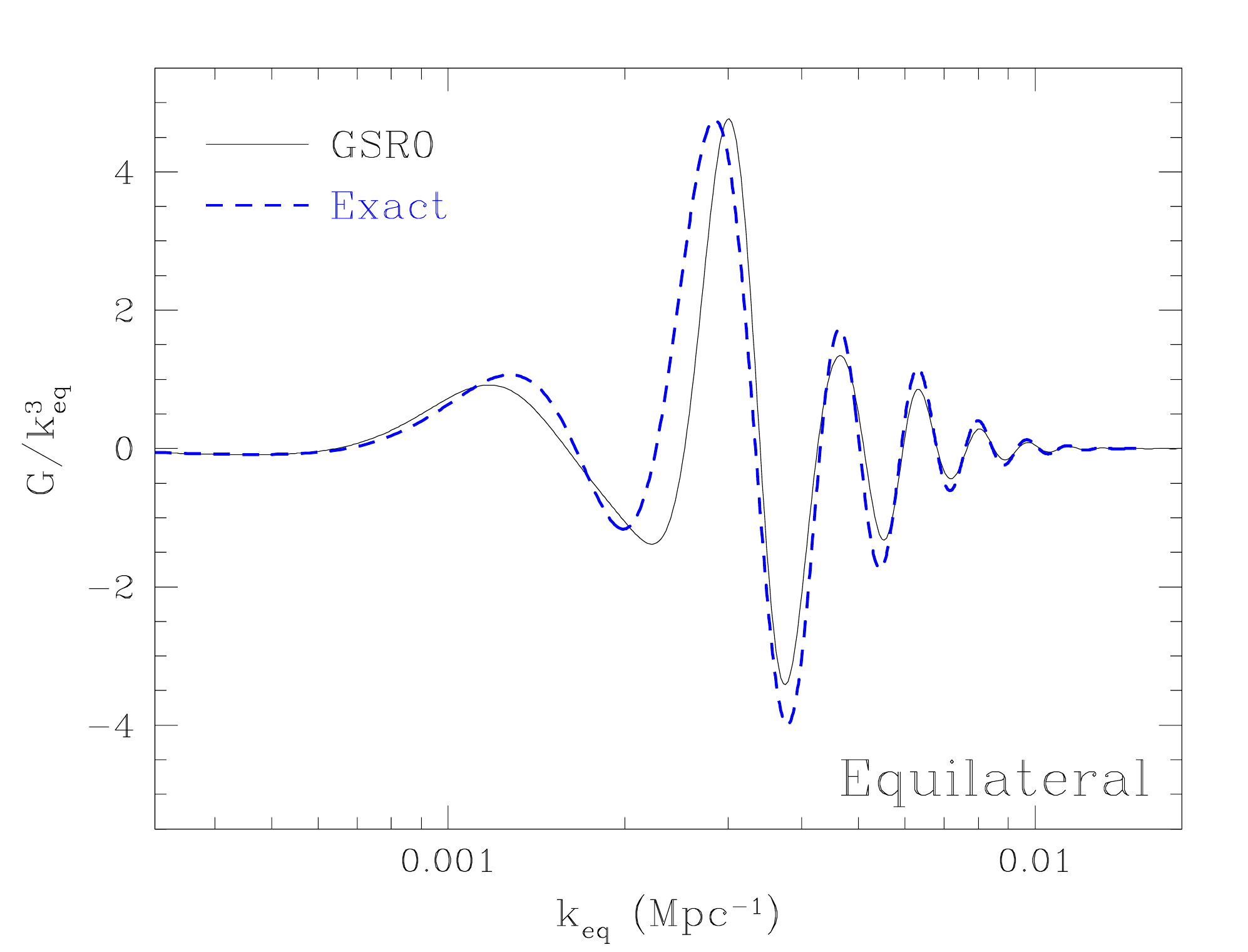, width=3.2in}}
\caption{\footnotesize Zeroth-order GSR approximation for equilateral configurations in the step potential model compared with exact results of evaluating 
Eq.~(\ref{eqn:bispectrum1st}).}
\label{fig:biapprox0equilateral}
\end{figure}

In the equilateral limit $k_1=k_2=k_3=k_{\rm eq}$ and
 Eqs.~(\ref{eqn:gsrlbi}) and (\ref{eqn:dimensionlessbi}) simplify  to
 \begin{align}\label{eqn:equibi}
 \frac{\mathcal{G}(k_{\rm eq}, k_{\rm eq},k_{\rm eq})}{k_{\rm eq}^3} &=
 {\Delta_{\cal R}^3(k_{\rm eq})\over 4 \tilde A_S^2} 
 \left[ -I_0- 6I_1 + 9 I_2 \right]_{K=3 k_{\rm eq}}. 
\end{align}
For equilateral triangles, the contribution from $I_0$ dominates the result (see Fig.~\ref{fig:integrals}) and
hence the comparison  with the numerical evaluation of Eq.~(\ref{eqn:bispectrum1st}) tests this aspect of the GSR approximation
(see Fig. \ref{fig:biapprox0equilateral}).  The zeroth-order GSR approximation 
captures the amplitude and phase of the oscillations fairly well.   The
largest deviations are at low $k$ for modes that were right on the horizon when
the inflaton crossed the feature in Fig.~\ref{fig:source}.  We show in Appendix \ref{app:GSR}
that these are associated with the neglect of first order terms involving other
combinations of the real and imaginary parts of the mode functions.

\subsection{Squeezed Limit}

\begin{figure}[t]
\centerline{\psfig{file=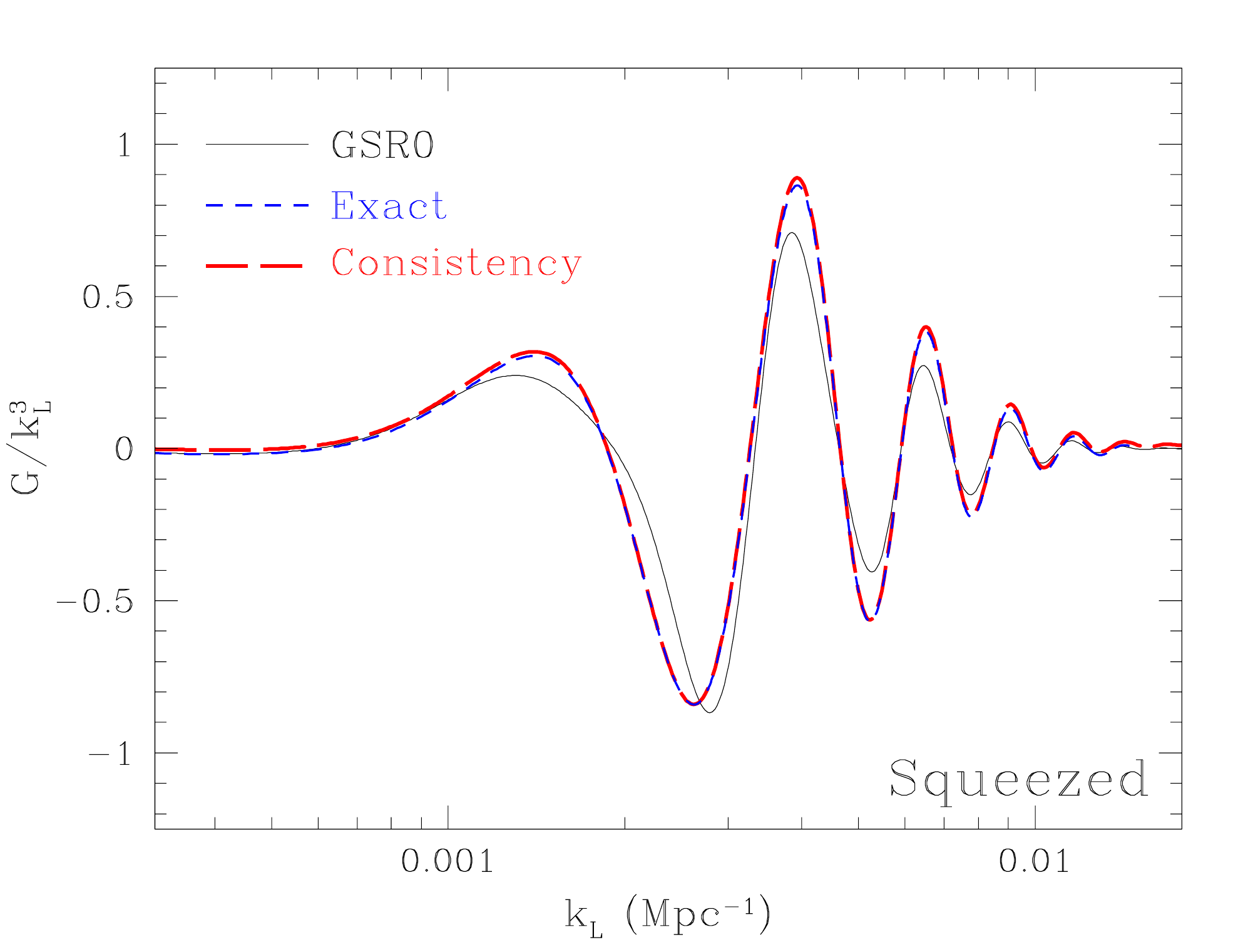, width=3.2in}}
\caption{\footnotesize Zeroth-order GSR approximation for squeezed bispectra 
$k_1=k_S \ll k_2 \approx k_3 = k_L$ in the step potential model compared with exact results of evaluating 
Eq.~(\ref{eqn:bispectrum1st}). Here $k_S=10^{-5}$ Mpc$^{-1}$.  For comparison
the prediction from the consistency relation with the slope of the curvature power
spectrum is also shown.}
\label{fig:biapprox0squeezed}
\end{figure}

For squeezed triangles  $k_S \equiv k_{1}\ll k_{2}\approx k_{3} \equiv k_L$
and  
 \begin{align}
 \frac{\mathcal{G}(k_S, k_L,k_L)}{k_{L}^3} =& 
 {\Delta_{\cal R}(k_S) 
\Delta^2_{\cal R}(k_L)\over 4 \tilde A_S^2} 
\left[ - 2I_1 + 4 I_2 \right]_{K= 2 k_L}.
\end{align}
Note that in this limit only $I_1$ and $I_2$ contribute 
and hence the comparison in Fig.~\ref{fig:biapprox0squeezed} tests a different
aspect of the GSR approximation.   In particular $I_1$ and $I_2$ carry the main impact
of the source correction discussed below Eq.~(\ref{eqn:Integrals}) since their windows carry superhorizon weight. 
In Appendix \ref{app:GSR}, we use these triangle configurations to develop and
test our approximation.

As a further check,
it is well known that the bispectrum of curvature fluctuations produced by an inflationary model with a single `clock' obeys a consistency relation which relates the squeezed limit of the bispectrum to the slope of the power spectrum \cite{Maldacena:2002vr}. The squeezed limit corresponds to one of the curvature fluctuations in the bispectrum having a much longer wavelength than the remaining two $k_{S}\ll k_{L}$.  In this limit, the consistency relation implies \cite{Creminelli:2004yq} 
\begin{equation}
 \frac{\mathcal{G}(k_S, k_L,k_L)}{k_{L}^3}  \approx 
-{\Delta_{\cal R}^2(k_L)\Delta_{\cal R}^2(k_S) \over 4\tilde A_S^2}
{d \ln \Delta_{\cal R}^2 \over d\ln k}\Big|_{k_L}.
\end{equation}
In addition to the comparison of our approximation to the numerical results in this limit, we use the numerically computed local slope to test the consistency relation itself. The result is plotted in Fig.~\ref{fig:biapprox0squeezed}, which shows excellent agreement between the numerically computed bispectrum in the squeezed limit, and the result obtained from the slope of the power spectrum and the consistency relation. There are small discernible differences away from the effects of the feature; however, we attribute these to our truncation of the action at Eq.~(\ref{eqn:truncatedaction}).

\subsection{Flat Limit}

\begin{figure}[t]
\centerline{\psfig{file=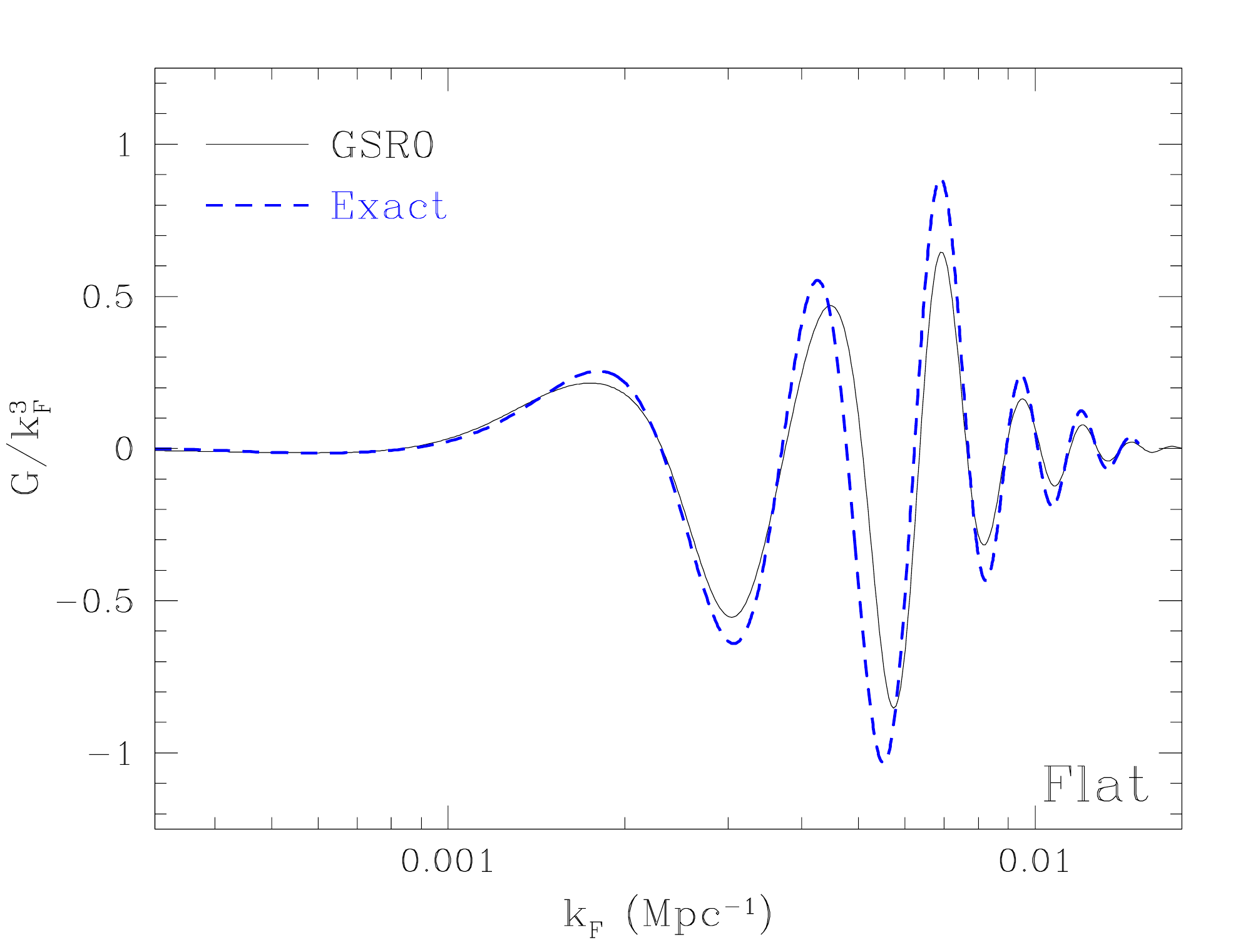, width=3.2in}}
\caption{\footnotesize Zeroth-order GSR approximation for flat  configurations 
with $k_1=k_F$ with $k_2=k_3=k_F/2$
 in the step potential model compared with exact results of evaluating 
Eq.~(\ref{eqn:bispectrum1st}). }
\label{fig:biapprox0flat}
\end{figure}

The final limit we consider is the flat limit, where $k_F = k_1 = 2 k_2 = 2 k_3$
and the wavevectors are co-linear.  Here
 \begin{eqnarray}
 \frac{\mathcal{G}(k_F, k_F/2,k_F/2)}{k_{F}^3} &=& 
 {\Delta_{\cal R}(k_F) 
\Delta^2_{\cal R}(k_F/2)\over 16 \tilde A_S^2} 
\\
&&\quad \times \left[ -I_0 - 7 I_1 + 12 I_2 \right]_{K=2 k_F}.\nonumber
\end{eqnarray}
and the approximation involves a combination of all three integrals.
In Fig.~\ref{fig:biapprox0flat}, we show that the approximation works comparably
well for flat triangles as equilateral and squeezed triangles.

\section{CMB Signal-to-Noise}
\label{sec:applications}

Given the step potential model that fits the glitches in the CMB power spectrum,
the corresponding features in the bispectrum described in the previous section 
\cite{Covi:2006ci,Hamann:2007pa,Mortonson:2009qv}
are a firm prediction \cite{Chen:2006xjb,Chen:2008wn}.   
What is less clear is to what extent they are observable.   Indeed that the model is designed to
fit low multipole  $\ell \sim 20-40$ glitches in the power spectrum implies that these
features will be strongly impacted by the cosmic variance of the
dominant Gaussian fluctuations \cite{Chen:2006xjb}.  On 
the other hand, bispectrum features generically extend to higher $k$ and hence $\ell$
than power spectra (see \S \ref{sec:GSRzero}).

An estimate of the signal-to-noise in the bispectrum for the step potential
has been hampered by the lack of a computationally efficient method for
estimating the curvature bispectrum.
Our zeroth-order GSR technique is ideal for 
these purposes as the bispectrum for any configuration can be simply formed
from three precomputed integrals in Eq.~(\ref{eqn:Integrals}).

\subsection{Cosmic Variance}

The temperature or angular bispectrum is defined as the three-point function 
of the spherical harmonic coefficients $a_{\ell m}$ of the temperature
anisotropy
\begin{equation}
B_{\ell_1 \ell_2 \ell_3} =\!\!\!\! \sum_{m_1 m_2 m_3} \left(
\begin{array}{ccc}
\ell_1 & \ell_2 & \ell_3 \\
m_1 & m_2 & m_3 
\end{array}
\right) \langle a_{\ell_1 m_1} a_{\ell_2 m_2} a_{\ell_3 m_3} 
\rangle .
\end{equation}
The cosmic variance of the Gaussian part of the field puts an irreducible
limit on the signal-to-noise ratio of 
\begin{equation}
\left( { S \over N}\right)^2 = \sum_{\ell_3 \ge \ell_2 \ge \ell_1} {B^2_{\ell_1\ell_2\ell_3} \over C_{\ell_1} C_{\ell_2} C_{\ell_3}
d_{\ell_1\ell_2\ell_3}}\, ,
\label{eqn:sn}
\end{equation}
where
\begin{equation}
d_{\ell_1 \ell_2\ell_3} = 
[1+ \delta_{\ell_1\ell_2} + \delta_{\ell_2\ell_3} + \delta_{\ell_3 \ell_1} +
2\delta_{\ell_1 \ell_2} \delta_{\ell_2 \ell_3}]
\end{equation}
accounts for permuted contractions of repeated $\ell$'s 
and the angular power spectrum is defined by
\begin{equation}
\langle a^*_{\ell m} a_{\ell ' m'} \rangle = \delta_{\ell \ell'} \delta_{m m'} C_\ell \,.
\end{equation}
Thus to evaluate the signal-to-noise in the angular bispectrum we require
not only a fast method for computing the curvature bispectrum but also
for computing angular bispectra from curvature bispectra.

\subsection{Approximations}

To obtain an order of magnitude estimate for the signal-to-noise ratio,
we seek only a crude computation of the angular bispectrum from the
curvature bispectrum.  We therefore take the flat-sky approach
and the Sachs-Wolfe limit for the temperature anisotropy.

 In the flat-sky approximation, the angular bispectrum is defined
 by the three-point function of the Fourier moments
 of the temperature field given by $a({\bf l})$
 \begin{equation}
\langle a({\bf l}_1) a({\bf l}_2) a({\bf l}_3) \rangle 
= (2\pi)^2 \delta({\bf l}_1+ {\bf l}_2+{\bf l}_3) B_{(\ell_1,\ell_2,\ell_3)} .
\end{equation}
 For $\ell_1$, $\ell_2$, $\ell_3 \gg 1$, it is related to the all-sky bispectra as
 \cite{Hu:2000ee}
 \begin{eqnarray}
B_{\ell_1 \ell_2 \ell_3} &=& \sqrt{ (2\ell_1+1)(2\ell_2+1)(2\ell_3+1)\over 4\pi} 
\left(
\begin{array}{ccc}
\ell_1 & \ell_2 & \ell_3 \\
0 & 0 & 0 
\end{array} \right) \nonumber\\
&& \times
B_{(\ell_1,\ell_2,\ell_3)}. 
\end{eqnarray}

Under the Sachs-Wolfe approximation, the temperature field as a function 
of angle $\hat{\bf n}$ on the sky is the projection of the curvature field onto
the sphere at the recombination distance $D$ from the observer
\begin{equation}
a(\hat{\bf n}) = -{1\over 5} \curv({\bf x}= D{\hat {\bf n}})
= -{1 \over 5} \int {d^3 k \over (2\pi)^3} \curv_{\bf k} e^{i {\bf k} \cdot D{\hat{\bf  n}}},
\end{equation}
so that 
\begin{eqnarray}
 B_{(\ell_1,\ell_2,\ell_3)}
&=&  -{2 \over 5^3 D^4}\int_{0}^\infty{ d k_{1 \parallel} \over 2\pi}
 \int_{-\infty}^\infty
{ d k_{2 \parallel} \over 2\pi} 
\nonumber\\
&& \times B_{\curv}(k_1,k_2,k_3) 
\end{eqnarray}
where
\begin{eqnarray}
{\bf k}_1 &=& ({\bf l}_{1}/D,k_{1\parallel}), \nonumber\\
{\bf k}_2 &=& ({\bf l}_{2}/D,k_{2\parallel}), \nonumber\\
{\bf k}_3 &=& -{\bf k}_{1}-{\bf k}_{2},
\end{eqnarray}
and $\parallel$ is the direction along the line of sight, orthogonal to the
plane of the sky.   Note that even in this approximation, the signal-to-noise in the 
angular bispectrum is a five dimensional sum over the curvature bispectrum.

For consistency, we also compute $C_\ell$ under the same flat-sky, Sachs-Wolfe
approximation for the cosmic variance in Eq.~(\ref{eqn:sn})
\begin{equation}
C_\ell = {1 \over 5^2 D^2} \int{ d k_{ \parallel} \over 2\pi} P_\curv({\bf k}=({\bf l}/D,k_{\parallel})).
\end{equation}

\begin{figure}[t]
\centerline{\psfig{file=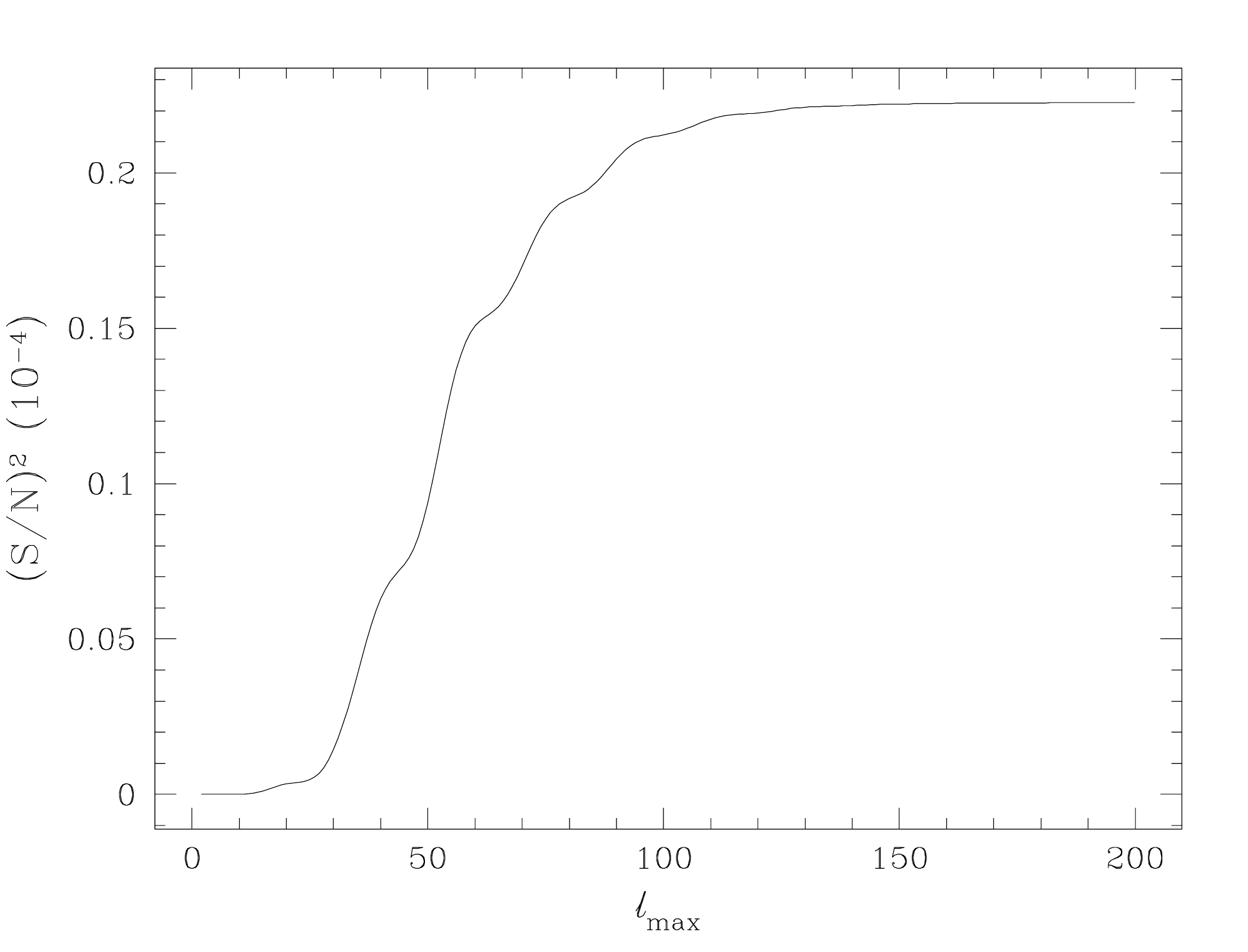, width=3.2in}}
\caption{\footnotesize Approximate cumulative signal-to-noise in the CMB angular bispectrum 
for the step model assuming cosmic variance limited temperature measurements out to a maximum multipole $\ell_{\rm max}$.   This model falls short of observable by more than $10^{4}$ in the number
of bispectrum triangles or $10^{2}$ in the amplitude of the bispectrum.}
\label{fig:sntot}
\end{figure}

In Fig.~\ref{fig:sntot} we show the cumulative signal-to-noise as a function of
the maximum $\ell$ in the sum of Eq.~(\ref{eqn:sn}) for the step model of \S \ref{sec:step}.   This model falls short of predicting observable effects by more than
$10^{4}$
in the number of bispectrum triangles or more than $10^{2}$ in the amplitude
of the bispectrum.  Consequently, the crudeness of our curvature to angular
bispectrum calculation is justified.

\section{Discussion}
\label{sec:discussion}

We have developed a computationally efficient technique based on the
generalized slow-roll approach for calculating the curvature bispectrum of
models with features in the inflaton potential.  
This technique allows all configurations of the bispectrum to be calculated
based on three precomputed integrals over the inflationary background 
reducing the dimensionality of the problem from three to one.

In this zeroth-order approximation, the accuracy is sufficient to capture the overall
amplitude and structure of the bispectra to typically better than 20\%.
We have also developed a first order approximation that that brings the typical
accuracy to better than several percent at the expense of raising the dimensionality
to two.   

The accuracy of the zeroth-order approximation more than suffices
to make an estimate of the observability of the bispectrum features in the
step potential model that best fits the WMAP power spectrum glitches at
$\ell \sim 20-40$.  We find that the bispectrum amplitude is more that a factor
of 100 too small to be observable in a cosmic variance limited measurement
of CMB temperature anisotropy.

We have also explicitly verified that the consistency relation holds for a model of inflation which violates slow roll due to a sharp downward step feature in its potential. We find excellent agreement through the region affected by the feature, while small disparities at smaller and larger wavenumber are consistent with our neglect of additional terms suppressed by $\mathcal{O}(\epsilon_H)$ in the cubic action.

Two key general insights arise from the comparison of the GSR approach to the
power spectrum and bispectrum.  First,  the dominant source of both is a similar combination of slow-roll parameters from the solution of the inflationary background.  
Second, the bispectrum is generated while modes were deeper within the horizon
compared with the power spectrum.  

The latter fact implies that features
in the bispectrum generically persist to higher wavenumber $k$ than the 
power spectrum and provides a template for constructing models with large
bispectrum features but small power spectrum features.   

Indeed 
recent work has revealed an intriguing `decoupling' limit under which large non-Gaussianity may be produced while the power spectrum remains largely unperturbed
 \cite{Flauger:2010ja, Leblond:2010yq}. The application of the techniques we develop here may be able to extend the region of validity for these scenarios into the region where gravity is not completely decoupled, merely very weakly coupled. 
We defer these considerations to a future work.

\acknowledgements

We thank  Eugene Lim for providing numerical data for comparison. This work was supported in part by the Kavli Institute for Cosmological Physics at the University of Chicago through grants NSF PHY-0114422 and NSF PHY-0551142 and an endowment from the Kavli Foundation and its founder Fred Kavli.   WH
 was additionally supported by U.S.~Dept.\ of Energy contract
 DE-FG02-90ER-40560 and the David and Lucile Packard Foundation. HVP is supported by Marie Curie grant MIRG-CT-2007-203314 from the European Commission, and by STFC and the Leverhulme Trust, and acknowledges the hospitality of the KICP.

\appendix

\section{Generalized Slow Roll}\label{app:GSR}

We begin in \ref{app:GSRreview} by reviewing 
 the generalized slow-roll (GSR) formalism \cite{Stewart:2001cd, Choe:2004zg, Dvorkin:2009ne}, collecting some important results and establishing our notation. 
 Employing the same methods used to correct the power spectrum
 in  \ref{app:GSRpower}, we derive the GSR expansion in for the 
 bispectrum in \ref{app:GSRbispectrum}.   We apply this expansion to the zeroth-order expansion 
 used for the main results of the paper in \ref{app:GSRzeroth}.   This expression
 involves corrections for the superhorizon modes based on a more accurate first order approach
 that we develop and test in \ref{app:GSRfirst}.  
 
  The treatment and notations used here mirror the presentation in \cite{Dvorkin:2009ne}. The reader is directed to these earlier works for further details.
For notational compactness we express functions of the three separate bispectrum
$k$ values as
\begin{equation}
F(k_i) \equiv F(k_1,k_2,k_3)
\end{equation}
throughout this Appendix.

\subsection{GSR Corrections to the Mode Functions}
\label{app:GSRreview}

Varying the action (\ref{eqn:quadaction}), the mode functions, $\mathcal{R}_{k}(\eta)$, satisfy the Mukhanov-Sasaki equation
\begin{align}
\frac{d^{2}\mathcal{R}_k}{dx^{2}}- \left( 1 - {f' \over f} \right) \frac{2}{x}\frac{d\mathcal{R}_k}{dx} + \mathcal{R}_k = 0,
\end{align}
where $x=k\eta$ and $f$ were defined in Eq.~(\ref{eqn:fdef}).

We seek to expand the mode functions in a perturbative series 
around their infinitely slow-roll or de Sitter form.   Since the comoving curvature ${\cal R}_k$ is undefined in
that limit, it is useful to change variables to 
\begin{equation}
y_i = \sqrt{k^3 \over 2\pi^2} {f \over x} \mathcal{R}_{k_i}.
\end{equation}
This variable $y_i$ also carries the interpretation of the inflaton field fluctuation
in the spatially flat gauge and is well defined in the de Sitter limit.  

With this correspondence it is easy to see that well inside the horizon $y_i$ is more immune
to features in the inflaton potential than ${\cal R}_k$ as it represents a free field.   The converse is true
outside the horizon.  The latter is the primary flaw in GSR that we seek to rectify
by enforcing constant curvature outside the horizon.  

As an aside, note that similar  superhorizon issues arise even for the computation of the bispectrum with
the exact ${\cal R}_k$.  Although ${\cal R}_k$ does not respond significantly to the feature on such scales, ${\cal R}_k'$ does and in
such a way to exactly cancel the behavior in the $\epsilon_H$, $\eta_H$ sources 
to keep the bispectrum constant in
Eq.~(\ref{eqn:bispectrum1st}).

We then have
\begin{align}\label{eqn:modeequation}
\frac{d^{2}y_i}{dx^{2}}+\left(1-\frac{2}{x^{2}}\right)y_i = \frac{g(\ln \eta)}{x^{2}}y_i,
\end{align}
where 
\begin{align}
g = \frac{f''-3f'}{f},
\end{align}
and primes denote derivatives with respect to $\ln\eta$.

The homogeneous Eq. (\ref{eqn:modeequation}) corresponds to exact de Sitter space ($H = {\rm const.}$) and has solutions
\begin{align}
y_{0}(x) = \left(1+\frac{i}{x}\right)e^{ix},
\end{align}
and $y_{0}^*(x)$ which depend in the same way  on $x=k\eta$ for all $k$. 
Given these solutions, we can use the Green function of the homogeneous operator to invert Eq.~(\ref{eqn:modeequation}),
\begin{align}\label{eqn:formal}
y_i(x) = y_{0}(x)+L(x,u) y_i(u),
\end{align}
with 
\begin{align}
L(x,u) y_i(u) =  -\int_{x}^{\infty}\frac{d u }{u^2}
g(\ln \tilde\eta)y_i(u)\Im[y^*_{0}(u)y_{0}(x)],
\end{align}
where $u = k\tilde\eta$.  Note that unlike $y_0$, $y_i$ is not the same
function of $x=k\eta$ for all $k$.

Then, assuming that the new solution is not too different from the de Sitter space result, we can employ the Born approximation and to solve the formal solution in Eq.~(\ref{eqn:formal}) iteratively
\begin{align}\label{eqn:iterativesol}
y_i(x) &=y_{0}(x) +L(x,u)y_{0}(u)
+L(x,w)L(w,u)y_{0}(u)+\ldots 
\end{align}
We will also use the notation
\begin{align}\nonumber
\WP(u)  = & -\frac{3}{u}\Im[y_{0}(u)]\Re[y_{0}(u)] \\\nonumber
 =& \frac{3\sin(2u)}{2u^{3}} - \frac{3\cos(2u)}{u^{2}}-\frac{3\sin(2u)}{2u},\\\nonumber
\XP(u) =& \frac{3}{u}\Re[y_{0}(u)]\Re[y_{0}(u)]\\\nonumber
 = &-\frac{3\cos(2u)}{2u^{3}} - \frac{3\sin(2u)}{u^{2}}+\frac{3\cos(2u)}{2u}\\
& +\frac{3}{2u^{3}}(1+u^2),
\end{align}
and
\begin{align}
\Im[y_{0}(u)]\Im[y_{0}(u)] = 1 +\frac{1}{u^{2}}-\frac{u}{3}\XP(u).
\end{align}

In the limit of small $u$, these window functions behave as $\lim_{u\rightarrow 0}\WP(u)  = 1$ and $\lim_{u\rightarrow 0}\XP(u)  = u^{3}/3$. 
\subsection{Power Spectrum Expansion}
\label{app:GSRpower}

The curvature power spectrum is given by
\begin{align}
\Delta_{\curv}^{2}(k) = \lim_{x\ll 1} \frac{x^{2}}{f^{2}}y_i(x)y_i^{*}(x) \,.
\end{align}
At zeroth-order in the mode function correction and first order in $g$ we obtain 
\cite{Stewart:2001cd}
\begin{align}\label{eqn:smallxfirstorder}
\lim_{x\rightarrow0}(xy_i) =&  i-\frac{i}{3}\int_{x}^{\infty}\frac{du}{u}\frac{x^{3}}{u^{3}}g(\ln\eta)\\\nonumber
&+\frac{i}{3}\int_{x}^{\infty}\frac{du}{u}\WP(u)g(\ln \eta)\\\nonumber
&-\frac{1}{3}\int_{x}^{\infty}\frac{du}{u}\XP(u)g(\ln \eta) + \mathcal{O}(x^{2}),\nonumber
\end{align}
(which corrects a sign typo for the third term in Eq.~(19) \cite{Dvorkin:2009ne})
where $x=k\eta_*$ and $u=k\eta$
and so
\begin{equation}
\Delta_{\curv}^{2}(k) = {1 \over f^2} \left[ 1 + {2 \over 3} {f'\over f} 
+ {1 \over 3} \int_{\eta_*}^\infty {d\eta \over \eta} \WP(k\eta) g(\ln \eta) \right],
\label{eqn:gsrs}
\end{equation} 
since the $\XP$ term contributes quadratically.

The problem with this expression is that the power spectrum depends on the 
arbitrary end point of the integration $\eta_*$.   The origin of this problem is that
the curvature computed in the GSR approximation is not guaranteed to be 
constant outside the horizon and constancy is only enforced order by order
in $g$ \cite{Dvorkin:2009ne}.    With large fluctuations, contributions that are formally higher order
in $g$ supply necessary corrections.   

To find these corrections, we can compute all of the first-order mode function corrections \cite{Choe:2004zg} and
retain the ones that contribute on superhorizon scales  \cite{Dvorkin:2009ne}
\begin{align}\label{eqn:GSRLfullpower}
\ln\Delta^{2}_{\rm GSR}(k) = G(\ln\eta_{*})+\int_{\eta_{*}}^{\infty}\frac{d\eta}{\eta}\WP(k\eta)G'(\ln\eta),
\end{align}
where the modified source is
\begin{align}
G'(\ln\eta) = \frac{2}{3}\left( g  - {f'^2 \over f^2} \right).
\end{align}
Formally $G'$ involves a first-order correction to the source since $f'/f = {\cal O}(g)$.
Note that for superhorizon modes when the inflaton crosses the feature $\WP(k\eta)
\rightarrow 1$ and the power spectrum no longer depends on the arbitrary 
end point.

We will follow this procedure to define our zeroth-order bispectrum formulation 
in Eq.~(\ref{eqn:gsrlbi}). 

\subsection{Bispectrum Expansion}
\label{app:GSRbispectrum}
In terms of the above notation, the bispectrum in Eq.~(\ref{eqn:bispectrum1st}) takes the form
\begin{align}\label{eqn:bispectrumall}
B_{\mathcal{R}}(k_{i}) & = \frac{(2\pi)^4\eta_*^3}{4k_{1}k_{2}k_{3}f^3}\Re\Big\{ iy_1(k_{1}\eta_*)y_2(k_{2}\eta_*)y_3(k_{3}\eta_*)\nonumber \\ & \int^{\infty}_{\eta_{*}}\frac{d\eta}{\eta}\gB(\ln\eta)
D_\eta [y_1^*(k_{1}\eta)y_2^*(k_{2}\eta)y_3^*(k_{3}\eta)]
\Big\}\nonumber\\
&+ {\rm Boundary\;\; Terms},
\end{align}
and the differential operator
\begin{equation}
D_\eta = \frac{d}{d\ln \eta}+3\left(1-\frac{f'}{f}\right).
\end{equation}
In the Eq.~(\ref{eqn:bispectrumall}), ``Boundary Terms'' refers to the second term in Eq.~(\ref{eqn:bispectrum1st}). While these terms are physically important, they represent a small correction and we defer their discussion to a subsection. 

Here the unmodified source function, which we will subsequently modify
in a similar prescription to $g \rightarrow 3 G'/2$, is
\begin{align}
\gB(\ln\eta) = \frac{( \epsilon_H - \eta_{H} )'}{f}.
\label{eqn:originalsource}
\end{align}
By analogy to the power spectrum window functions, let us define  the window functions
\begin{align}\nonumber
 W_{B}(x_{i}) &\equiv \Re[y_1(x_{1})y_2(x_{2})y_3(x_{3})] ,\\
X_{B}(x_{i}) &\equiv \Im[y_1(x_{1})y_2(x_{2})y_3(x_{3})] .
\end{align}
For the zeroth-order mode functions, $y_i \rightarrow y_0$, $W_B \rightarrow W_B^0$,
$X_B \rightarrow X_B^0$
\begin{eqnarray}
W_B^0(x_i) &=& {1 \over x_1 x_2 x_3} \Big[
(-X+ x_1 x_2 x_3) \cos X \nonumber\\&&\quad + (1  - x_2 x_3 - x_1 x_2 - x_1 x_3)\sin X
\Big], \nonumber\\
X_B^0(x_i) &=&  -{1 \over x_1 x_2 x_3} \Big[(1 - x_2 x_3 - x_1 x_2 - x_1 x_3) \cos X
\nonumber\\&&\quad  + (X  - x_1 x_2 x_3)\sin X\Big],
\label{eqn:WXB0}
\end{eqnarray}
where $X = x_1+x_2+x_3$, and is not to be confused with the window function $X(k \eta)$.  Note that the trigonometric functions depend only 
on the perimeter of the triangle $K$ and that the prefactors are pure powers of $k_i$.
This is a critical simplification achieved by  using zeroth-order mode functions.

As we shall see, their superhorizon limit  will be particularly important
\begin{align}\label{eqn:window0limit}
\lim_{k_{i}\rightarrow 0}W^{0}_{B}(k_{i}\eta) = & \frac{k_{1}^{3}+k_{2}^{3}+k_{3}^{3}}{3k_{1}k_{2}k_{3}} + \mathcal{O}(k_i^{2}\eta^{2}),\\\nonumber
\lim_{k_{i}\rightarrow 0}X^{0}_{B}(k_{i}\eta) =& -\frac{1}{k_{1}k_{2}k_{3}\eta^{3}}+\frac{1}{2}\frac{k_{1}^{2}+k_{2}^{2}+k_{3}^{2}}{k_{1}k_{2}k_{3}\eta}
 + \mathcal{O}(k_i\eta).
\end{align}
To an arbitrary order, the bispectrum can be written
\begin{align}
B_{\mathcal{R}}(k_{i}) &\approx \frac{(2\pi)^4}{4k_{1}k_{2}k_{3}}\frac{\eta_*^3}{f^{3}} 
  \Re\Big\{  i [ W_{B}(k_{i}\eta_*) +iX_{B}(k_{i}\eta_*)]\nonumber \\&\quad\times\int^{\infty}_{\eta_{*}}\frac{d\eta}{\eta}\gB(\ln\eta) D_\eta  \left[W_{B}(k_i\eta)-iX_{B}(k_{i}\eta)\right]
\Big\}. 
\end{align}

Specifically, using this notation we can write the zeroth and first-order bispectrum as
\begin{align}\label{eqn:fullbi1storder}
B_{\mathcal{R}}(k_{i}) & \approx \frac{(2\pi)^4}{4k_{1} k_{2} k_{3} }
 \frac{\eta_*^3}{f^{3}} \int^{\infty}_{\eta_{*}}\frac{d\eta}{\eta}\gB(\ln\eta) D_\eta
  \big\{ 
 \\
 &\quad W^{0}_{B}(k_i\eta_*)   X^{0}_{B}(k_{i}\eta) (1-\Re[\mathcal{C}(k_i,\eta)-\mathcal{C}(k_i,\eta_{*})]) \nonumber\\
& -X^{0}_{B}(k_i\eta_*) W^{0}_{B}(k_i\eta) (1-\Re[\mathcal{C}(k_i,\eta)-\mathcal{C}(k_i,\eta_{*})])\nonumber\\
&  + W^{0}_{B}(k_i\eta_*)W^{0}_{B}(k_i\eta) \, \Im[\mathcal{C}(k_i,\eta_{*}) -\mathcal{C}(k_i,\eta)] \nonumber\\
&
+X^{0}_{B}(k_i\eta_*) X^{0}_{B}(k_i\eta)\,  \Im[\mathcal{C}(k_i ,\eta_{*})-\mathcal{C}(k_i ,\eta)] \big\}, \nonumber
\end{align}
where
\begin{align}
\mathcal{C}(k_i,\eta) = \sum_{j = 1}^{3}\int^{\infty}_{\eta}{d \tilde\eta \over \tilde\eta}\frac{g(\ln \tilde\eta)}{k_{j}\tilde\eta}\frac{y_{0}(k_j\tilde\eta)}{y_{0}(k_{j}\eta)}\Im[y^{*}_{0}(k_{j}\tilde\eta) y_{0}(k_{j}\eta)].
\end{align}
The fact that $\mathcal{C}$ depends on the three $k$'s presents the main obstacle to simplifying the first-order expressions.

\begin{figure}[t]
\centerline{\psfig{file=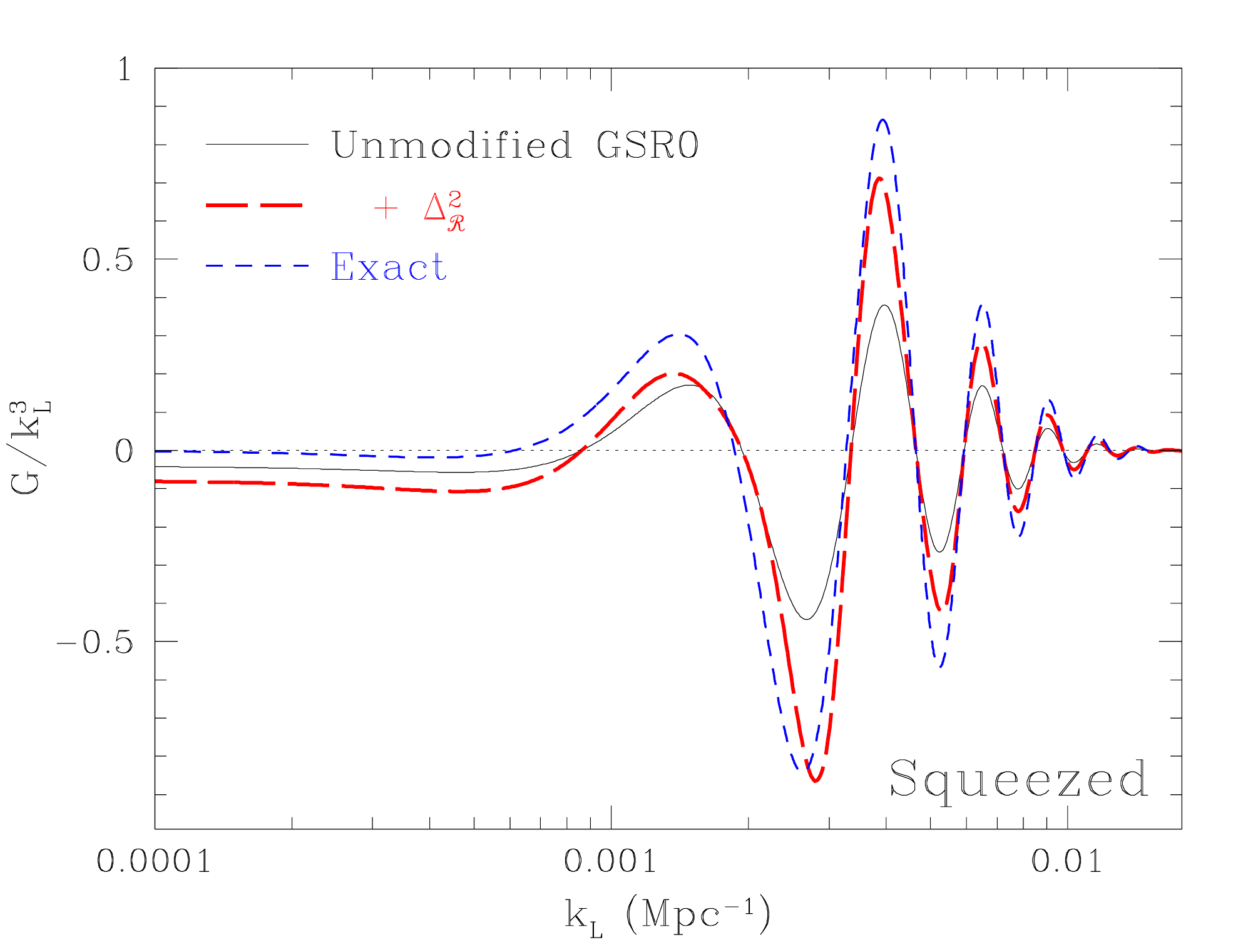, width=3.2in}}
\caption{\footnotesize Unmodified source with zeroth-order GSR approximation for
squeezed triangles as in Fig.~\ref{fig:biapprox0squeezed}. Solid
lines denote unmodified form (\ref{eqn:gsrsbi}).  Red dashed lines represent the replacement of the external $1/f$ with $\Delta_{\cal R}$ in Eq.~(\ref{eqn:extmodes}). Here $\eta_* = 1$Mpc.}
\label{fig:unmod}
\end{figure}

\subsection{Zeroth-Order Bispectrum}
\label{app:GSRzeroth}

We define the order of our approximation by whether the mode functions used
in the construction are the zeroth-order de Sitter $y_0$ or an iterative expansion.
Thus we allow ourselves the freedom to add 
higher order corrections to the source function but not the mode function.
This operational definition, rather than order counting in $g$  is motivated by
simple form of the bispectrum that results.   With only zeroth-order mode functions,
all triangles can be expressed in terms of single integrals that depend only on
the perimeter of the triangle.  

\subsubsection{Unmodified Source}

We begin with the zeroth-order approximation using the original source $\gB$ in
Eq.~(\ref{eqn:originalsource}). This form is consistently first order in the
deviations from slow roll and of the terms in Eq.~(\ref{eqn:fullbi1storder}) only the 
$X_B^0(k_i\eta_*)W_B^0(k_i\eta)$ leading order term survives
\begin{align}\!\!\!\!\!\!\!
B_{\curv}(k_i)  &\approx  {(2 \pi)^4  \over 4k_1^2 k_2^2 k_3^2} {1\over 
f^3}\Big[\int^{\infty}_{\eta_{*}}\frac{d\eta}{\eta}\gB(\ln\eta) ( W_B^0{}'+3  W_B^0)\Big].
\label{eqn:gsrsbi}
\end{align}
In Fig.~\ref{fig:unmod}, we compare this approximation with the exact result
for squeezed triangles.
There are two obvious flaws in this expression.  First the external $f = f(\ln\eta_*)$
factors depend on the arbitrary end point.   Second, in the superhorizon limit
$W_B^0$ goes to a constant defined in Eq.~(\ref{eqn:window0limit}).  Thus, analogous to the behavior of the power spectrum in the same approximation, the feature is imprinted on superhorizon modes.

\subsubsection{Source Modification}

Just as in the power spectrum case, we can fix these problems by examining
the first-order mode function corrections in the superhorizon limit.  
We start from the full first-order expression Eq.~(\ref{eqn:fullbi1storder}) and extract a factor of 
\begin{equation}
1-\Re[\mathcal{C}(k_i,\eta_{*})]  = 1+  \frac{f'}{f} + \sum_j \int^{\infty}_{\eta_{*}}\frac{d\eta}{\eta} {g(\ln\eta) \over 3} \WP(k_j\eta),
\end{equation}
where we have evaluated the expression in the limit $k_{i}\eta_{*}\ll1$.
We have also dropped higher order terms.

Combined with the factors of $1/f$ we recognize these factors as the
GSR expression (\ref{eqn:gsrs}) for the power spectrum.
Our prescription therefore is to replace these terms with the
exact power spectrum
\begin{align}\label{eqn:extmodes}
\frac{1}{f^{3}}(1-\Re[\mathcal{C}(k_{i},\eta_{*})]) = \Delta_{\cal R}(k_{1})\Delta_{\cal R}(k_{2})\Delta_{\cal R}(k_{3}).
\end{align}
  This fixes the problem of the
external $f$ factors in Eq.~(\ref{eqn:gsrsbi}).  In Fig.~\ref{fig:unmod}, we show
the impact of applying this correction.   The low $k$ features in particular 
are modulated
and enhanced by the external power spectra but the superhorizon problem 
remains.

 We then look at the superhorizon limit of the rest of the first-order corrections. 
 Since $W_{B}$ scales as a constant in the superhorizon limit, we need only keep the constant parts of the expansion of $\Re[\mathcal{C}(k_{i}, \eta)]$
\begin{align}
\lim_{k_i\rightarrow 0}  \Re[\mathcal{C}(k_i,\eta)]& = \sum_{j = 1}^{3}\Big[ 
\frac{1}{3}\int_{x_j}^{\infty}\frac{d\tilde{\eta}}{\tilde{\eta}}\frac{\eta^3}{\tilde{\eta}^{3}}g(\ln \tilde{\eta} )\\ & - \frac{1}{3}\int_{x_{j}}^{\infty}\frac{d\tilde{\eta}}{\tilde{\eta}}\WP(k_{j}\tilde{\eta})g(\ln \tilde{\eta})+\mathcal{O}(k_j^2\eta^{2}) \Big].\nonumber
\end{align}
On the other hand, $X_{B}$ behaves as $k_{i}^{-3}$, so we need to keep up to ${\cal O}(k_{i}^{3})$ parts of $\Im[\mathcal{C}(k_{i}, \eta)]$:
\begin{eqnarray}\nonumber
\!\!
\lim_{k_i \rightarrow 0}  \Im[\mathcal{C}(k_i,\eta)]& = & \sum_{j = 1}^{3}\Big[ -
\frac{1}{3}\int_{\eta}^{\infty}\frac{d\tilde{\eta}}{\tilde{\eta}}\XP(k_{j}\tilde{\eta})g(\ln \tilde{\eta}) \\ \nonumber
&& + \frac{2( k_{j}\eta)^{3}}{9}\int_{\eta}^{\infty}\frac{d\tilde{\eta}}{\tilde{\eta}}\WP(k_{j}\tilde{\eta})g(\ln \tilde{\eta}) \\\nonumber
&&- \frac{(k_{j}\eta)^{3}}{9}\int_{\eta}^{\infty}\frac{d\tilde{\eta}}{\tilde{\eta}}\frac{\eta^{3}}{\tilde{\eta}^3}g(\ln \tilde{\eta}) \\ &&+\mathcal{O}(k_j^5 \eta^{2}\tilde{\eta}^{3})\Big].
\end{eqnarray}
Then, working to order $\mathcal{O}(\gB {f'}/{f})$
\begin{align}\label{eqn:superHlimit}
B_{\mathcal{R}}(k_{i})&\approx {3 (2 \pi)^4  \over 4 k_1^2 k_2^2 k_3^2}
\Delta_{\cal R}(k_{1})\Delta_{\cal R}(k_{2})\Delta_{\cal R}(k_{3})  \frac{k_{1}^{3}+k_{2}^{3}+k_{3}^{3}}{3k_{1}k_{2}k_{3}}\nonumber\\ &\quad \times
\int^{\infty}_{\eta_{*}}\frac{d\eta}{\eta}\gB\Bigg[1-\int^{\infty}_{\eta} \frac{d\tilde{\eta}}{\tilde{\eta}}\frac{f'}{f}\Bigg].
\end{align}
Now, to a good approximation,
\begin{align}
\gB = \frac{( \epsilon_H-  \eta_{H})'}{f} \approx -\frac{1}{f}\left(\frac{f'}{f}\right)'.
\end{align}
With this approximation for the source, integrating the second line of Eq.~(\ref{eqn:superHlimit}) by parts one obtains
\begin{align}\label{eqn:sourceintbyparts}
-\int^{\infty}_{\eta_{*}}\frac{d\eta}{\eta}\left(\frac{1}{f}\frac{f'}{f}\right)'-\frac{1}{f}\frac{f'}{f}\int^{\infty}_{\eta_{*}} \frac{d\eta}{\eta}\frac{f'}{f}. 
\end{align}
Dropping the slow-roll suppressed second term, we can account for this first-order effect on superhorizon scales by the replacement
\begin{align}\label{eqn:sourcereplacement}
\gB \approx -\frac{1}{f}\left(\frac{f'}{f}\right)' 
\rightarrow   -\left(\frac{1}{f}\frac{f'}{f}\right)' 
\approx  \left( \frac{ \epsilon_H-  \eta_{H}}{f} \right)' \equiv G_B'.
\end{align}
In the superhorizon limit, the contribution of the first term in Eq.~(\ref{eqn:bispectrum1st}) to the bispectrum then reduces to the integral of a total derivative of slow-roll parameters, which  are small at late times, once the inflaton has settled back onto the slow-roll attractor. 

To summarize, our zeroth-order approximation consists of the replacement of the source by Eq.~(\ref{eqn:sourcereplacement}) and the replacement of the external factors using Eq.~(\ref{eqn:extmodes}) while using the de Sitter mode functions everywhere else. These replacements lead to the expression in Eq.~(\ref{eqn:gsrlbi}) with
the help of the explicit form for the $W_B^0$ window in Eq.~(\ref{eqn:WXB0}). 
In particular note that the $I_0$ term arises from the $W_B^0{}'$ term in 
Eq.~(\ref{eqn:gsrsbi}).

\subsubsection{Boundary Terms}

While small, the boundary term in Eq.~(\ref{eqn:bispectrum1st}) hitherto omitted
plays an important role in ensuring that the bispectrum becomes independent of time on superhorizon scales.

The contribution from the boundary term can be obtained from the results of the previous section with the replacement
\begin{align}
\int^{\infty}_{\eta_{*}}\frac{d\eta}{\eta}\gB(\ln\eta) \rightarrow \left.\frac{(\epsilon_H - \eta_H)}{f}\right|_{\eta = \eta_{*}},
\end{align}
together the replacement of all $\eta$ by $\eta_{*}$. Using this replacement on Eq.~(\ref{eqn:superHlimit}) yields for the boundary term
\begin{align}\label{eqn:superHlimitboundary}
B^{\rm BT}_{\mathcal{R}}(k_{i})&= {3 (2 \pi)^4  \over 4 k_1^2 k_2^2 k_3^2}
\Delta_{\cal R}(k_{1})\Delta_{\cal R}(k_{2})\Delta_{\cal R}(k_{3})  \frac{k_{1}^{3}+k_{2}^{3}+k_{3}^{3}}{3k_{1}k_{2}k_{3}}\nonumber\\ &\quad \times\left( \frac{ \epsilon_H-  \eta_{H}}{f} \right)\Bigg[1-\int^{\infty}_{\eta_*} \frac{d{\eta}}{\eta}\frac{f'}{f}\Bigg].
\end{align}
Notice that, when added to Eq.~(\ref{eqn:superHlimit}) using Eq.~(\ref{eqn:sourceintbyparts}), one obtains perfect cancellation leaving a time independent result.  Since the integral term in Eq.~(\ref{eqn:superHlimitboundary}) exactly cancels the second term in (\ref{eqn:sourceintbyparts}) we choose to omit both in 
 practice.   This brings the net boundary term to the $G_{B}(\ln\eta_{*})$ contribution to Eq.~(\ref{eqn:gsrlbi}).
 
 We also point out here that, while for convenience we have evaluated our expressions at a time when the inflaton has passed the feature and is back on its slow-roll attractor, our results are not restricted to this regime. As we pointed out in Section \ref{sec:methods}, the cancellation we are enforcing is exact and thus truly independent of the time at which the bispectrum is evaluated, even if this time is chosen to be when the slow-roll parameters are not small.

\subsection{First-Order Bispectrum}
\label{app:GSRfirst}

We can further compute to first order in the mode function correction on all scales.  
Note that the first order corrections represent a calculation of the bispectrum
to second order in the slow-roll parameters. Additionally, note that the boundary terms are always suppressed by slow-roll parameters evaluated at times well after the inflaton has traversed the feature which means that they are computationally irrelevant.

Neglecting boundary terms we obtain at first order in the GSR approximation, the full bispectrum
\begin{align}\nonumber
 B_{\curv}(k_i) 
=& {(2\pi)^4 \over 4} {\Delta_{\rm GSR}(k_{1}) \over k_1^2}{ \Delta_{\rm GSR}(k_{i}) \over k_2^2}{\Delta_{\rm GSR}(k_{3}) \over k_3^2}\\ & \times
  \int^\infty_{\eta_*} {d\eta\over\eta}\gB(\ln\eta) [ U_0 +
U_{1A} +U_{1B}+U_{1C}
&\nonumber\\
& \qquad+U_{1D}+U_{1E}](k_i\eta) ,
\end{align}
where
{\allowdisplaybreaks[4]
\begin{align}
U_0(k_{i}\eta)  =&  \left(\frac{d}{d\ln\eta}+3\right)\Re[y_{0}(k_{1}\eta)y_{0}(k_{2}\eta)y_{0}(k_{3}\eta)],
 \nonumber\\
 %
 \nonumber
 U_{1A}(k_{i}\eta) =& {1\over 2}  \int_{\eta_*}^\infty {d\tilde{\eta} \over \tilde{\eta}} G'(\ln\tilde{\eta}) \XP(k_{3}\tilde\eta)  \left(\frac{d}{d\ln\eta}+3\right)\\ \nonumber& \times
\Im[y_{0}(k_{1}\eta)y_{0}(k_{2}\eta) [y^{*}_{0}(k_{3}\eta)+y_{0}(k_3\eta)]]\\ & \qquad+{\rm cyc.}, \nonumber\\
%
%
\nonumber
U_{1B}(k_{i}\eta)=& -{1\over 2} \int_{\eta}^\infty {d\tilde{\eta} \over \tilde{\eta}} G'(\ln\tilde{\eta}) \WP(k_{3}\tilde\eta)
\\ &\times
 \left(\frac{d}{d\ln\eta}+3\right)\Re[y_{0}(k_{1}\eta)y_{0}(k_{2}\eta)y^{*}_{0}(k_{3}\eta)] \nonumber\\&\quad + {\rm cyc.},\nonumber\\
 %
\nonumber
U_{1C}(k_i\eta) = &
- {1\over 2} \int_{\eta_*}^\eta {d\tilde{\eta} \over \tilde{\eta}} G'(\ln\tilde{\eta})  \XP(k_{3}\tilde\eta)\\\nonumber &\times
   \left(\frac{d}{d\ln\eta}+3\right)\Im[y_{0}(k_{1}\eta)y_{0}(k_{2}\eta)y^{*}_{0}(k_{3}\eta)]\\ & +{\rm cyc.},\nonumber\\
%
\nonumber
U_{1D}(k_i\eta) =& - {3\over 4} \int_{\eta}^\infty {d\tilde{\eta} \over \tilde{\eta}} G'(\ln\tilde{\eta}) \left({1 \over k_{3}\tilde\eta} + {1 \over (k_{3}\tilde\eta)^3}\right)\\\nonumber &
\left(\frac{d}{d\ln\eta}+3\right)\Im[y_{0}(k_{1}\eta)y_{0}(k_{2}\eta)\\ &
[y^{*}_{0}(k_{3}\eta)+y_{0}(k_{3}\eta)]] + {\rm cyc.},\nonumber\\
 %
 \nonumber
 U_{1E}(k_i\eta) =& - 3
\Re[y_{0}(k_{1}\eta)y_{0}(k_{2}\eta)y_{0}(k_{3}\eta)]\\ &\times {f' \over f} \left[ 1 - {1\over 2 \gB} \left( {f' \over f} \right)^2 \right],
\end{align} 
where cyc. denotes the 2 additional cyclic permutations of the $k$ indices.
}

In order to fix a small second order correction to the superhorizon results we
have replaced $g \rightarrow 3G'/2$ in the the first order expressions and likewise
replace the source term
\begin{equation}
{f' \over f} \rightarrow {f' \over f} \left[ 1 - {1\over 2 \gB} \left( {f' \over f} \right)^2 \right] 
\end{equation}
in $U_{1E}$.  Furthermore for consistency, we use the GSR power spectrum
approximation (\ref{eqn:GSRLfullpower})  for the external terms.

\begin{figure}[t]
\centerline{\psfig{file=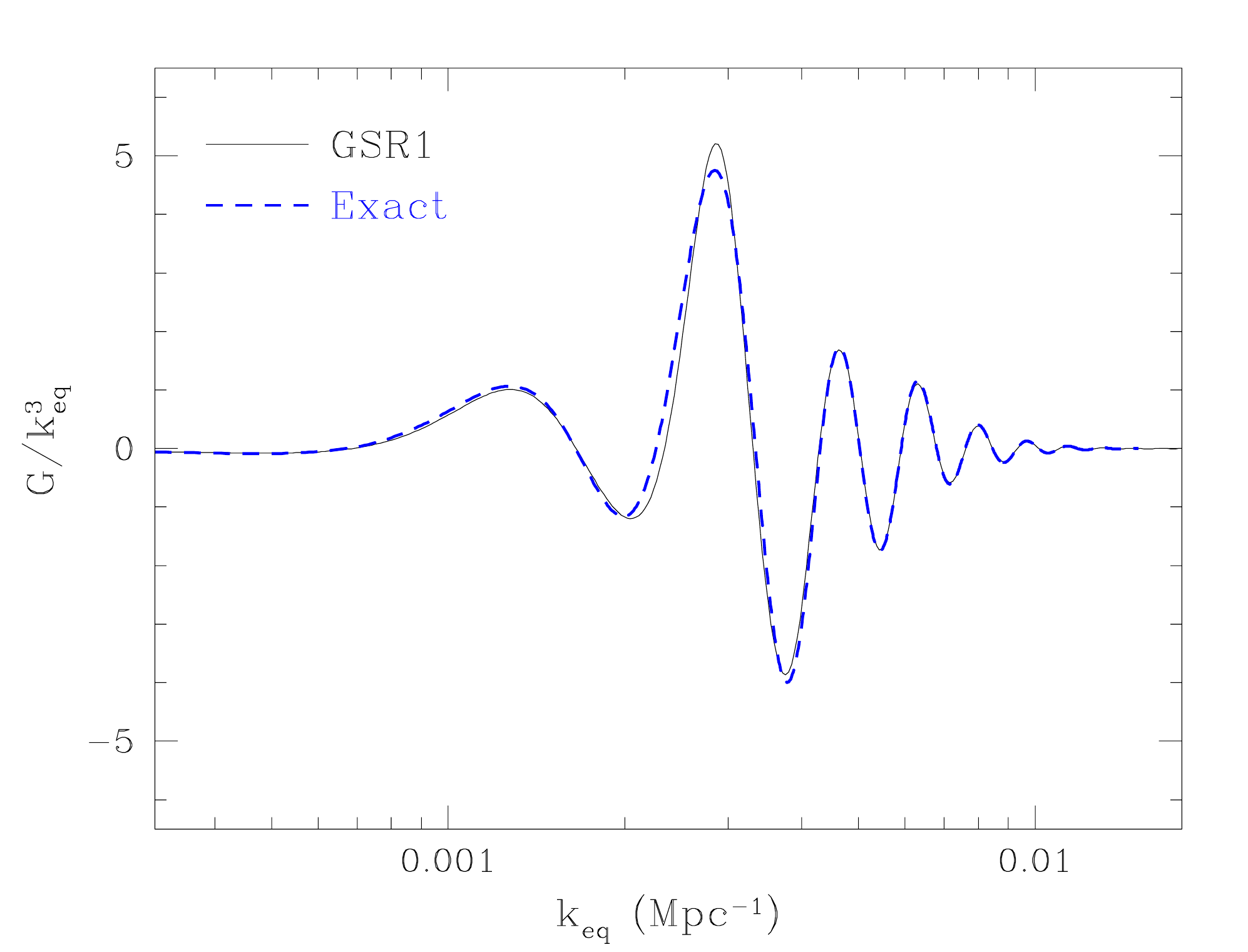, width=3.2in}}
\caption{\footnotesize First order GSR mode function approximation for the step model compared with the exact solution.}
\label{fig:biapprox1}
\end{figure}

For the case of the equilateral configurations, this expression simplifies 
considerably and are practical to evaluate
\begin{align}
 B_{\curv}(k,k,k) 
&= {(2\pi)^4 \Delta_{\rm GSR}^3(k) \over 4 k^6} 
\int_{\eta_*}^\infty {d\eta\over\eta}\gB(\ln\eta) [ U_0^{\rm eq} \nonumber\\
& \quad + U_{1A}^{\rm eq}  + U_{1B}^{\rm eq} + U_{1C}^{\rm eq} + U_{1D}^{\rm eq} +U_{1E}^{\rm eq} ](k\eta),
\end{align}
where
{\allowdisplaybreaks[4]
\begin{eqnarray}
U_0^{\rm eq}(v) &=&  -6 \cos 3v + {3(1-v^2) \sin 3v \over v},
 \nonumber \\
 U_{1A}^{\rm eq}(v) &=& {3\over 2v}
\big[ (3 + v^2) \cos v + 3 ( v^2-1) \cos 3 v 
\nonumber\\ && \quad + 
 2 v (\sin v - 3 \sin3 v) \big]
 \nonumber\\
 &&\times
 \int_{\eta_*}^\infty {d\tilde{\eta} \over \tilde{\eta}} G'(\ln\tilde{\eta}) \XP(k\tilde{\eta})  ,
\nonumber \\
U_{1B}^{\rm eq}(v) &=& {3\over 2}
\left[{(3+v^2)\sin v \over v}-2\cos v\right] \nonumber\\
&&\quad\times
  \int_{\eta}^\infty {d\tilde{\eta} \over \tilde{\eta}} G'(\ln\tilde{\eta}) \WP(k\tilde{\eta}),
 \nonumber \\
 U_{1C}^{\rm eq}(v) & =&
- {3\over 2} \left[ \left( {3\over v}+v\right) \cos v + 2 \sin v\right] \nonumber\\
&&\quad\times
\int_{\eta_*}^\eta {d\tilde{\eta} \over \tilde{\eta}} G'(\ln\tilde{\eta})  \XP(k\tilde{\eta})  ,
 \nonumber \\
U_{1D}^{\rm eq}(v) &=& - {9\over 4v}
\big[ (3 + v^2) \cos v + 3 ( v^2-1) \cos 3 v 
\nonumber\\ && \quad + 
 2 v (\sin v - 3 \sin3 v) \big]
\nonumber\\ && \quad \times
  \int_{\eta}^\infty {d\tilde{\eta} \over \tilde{\eta}} G'(\ln\tilde{\eta}) \left({1 \over \tilde v} + {1 \over \tilde v^3}\right),
 \nonumber \\
 U_{1E}^{\rm eq}(v) &=& - 
 {3\over v^3} \left[v (v^2-3) \cos 3 v + (1 - 3 v^2) \sin 3 v \right] 
 \nonumber\\ && \quad \times
{f' \over f} \left[ 1 - {1\over 2 \gB} \left( {f' \over f} \right)^2 \right].
\end{eqnarray}
}

We show the result of computing these terms for the step model of \S \ref{sec:step} in
Fig.~\ref{fig:biapprox1}. Note that most of the error in the zeroth-order approximation is corrected by the first order expression.  
The dominant correction is from $U_{1A}^{\rm eq}$ and involves the product of two independent integrals.  Since its 
 computation is no more intensive than the zeroth-order expression,  it may
be simultaneously computed as a monitor of the accuracy of the zeroth-order
expression.  Note that the integral of $G' \XP$ is the same as the one that monitors the accuracy of the power spectrum approximation \cite{DvoHu10a}.

\begin{figure}[t]
\centerline{\psfig{file=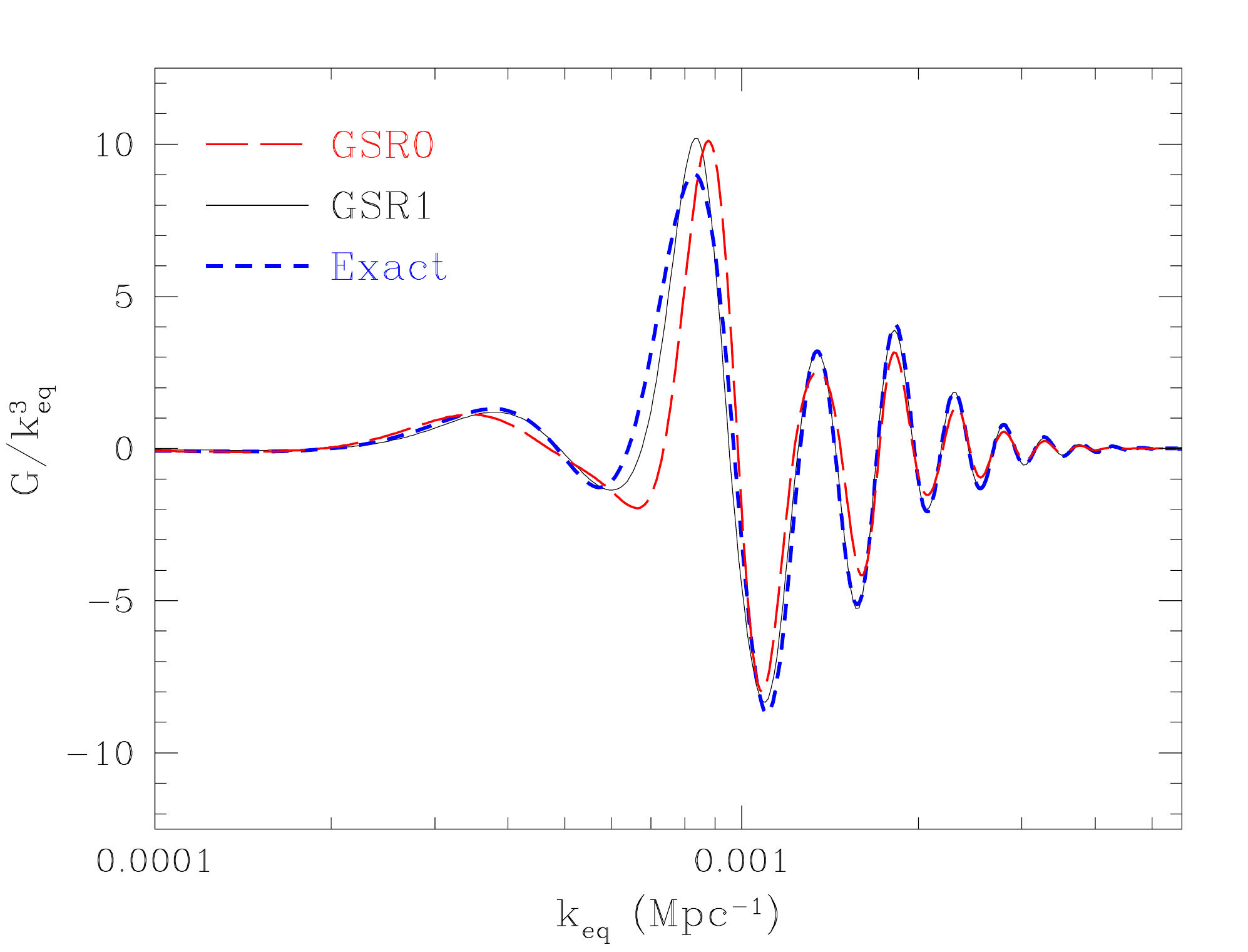, width=3.2in}}
\caption{\footnotesize Bispectrum in the step model of  Ref.~\cite{Chen:2006xjb,Chen:2008wn}
 and $\tilde A_S = 4.415 \times 10^{-10}$ (see text).   Our approximations work
equally well for models with larger bispectrum features.}
\label{fig:eugene}
\end{figure}

This first order 
mode function correction breaks the property that all triangles depend only
on integrals involving the perimeter $K$.   Thus though more accurate it
is of more limited utility for fast computations in that the six integrals involved are a function of
two variables: the perimeter $K$ and the wavenumber of the corrected mode function $k_3$.
Nonetheless it still reduces the complexity of the exact computation which requires
all three $k_i$ as well as the exact mode functions for each individual $k_i$.

\smallskip
\section{Comparison to Previous Work}
\label{app:eugene}
Previous calculations of the step bispectrum employed the parameters of: $\{m,c,d,\phi_{s}\} = \{3\times10^{-6}, 1.8\times 10^{-3}, 0.022, 14.84 \}$ 
\cite{Chen:2006xjb,Chen:2008wn}.
 There are three important differences between this model
and the one defined in \S \ref{sec:step}.  

First due to an error in the
setting of the initial conditions in Ref.~\cite{Covi:2006ci,Hamann:2007pa} the location of the feature $\phi_s$ is shifted and the
corresponding features in the power spectrum and bispectrum appear
at lower $k$.  Secondly, the maximum likelihood fit to WMAP5 data prefer
a slightly larger width $d$ compared with WMAP3 data.  
 Since the bispectrum is particularly sensitive
to the width $d$, its features appear more prominent and persist to higher $k$. 
Finally Ref.~\cite{Chen:2006xjb,Chen:2008wn}
arbitrarily set the parameter $m$ whereas Ref.~\cite{Covi:2006ci,Hamann:2007pa}
normalized to WMAP3.   As a consequence, we take $\tilde A_S = 4.415 \times 10^{-10}$ to be compatible with this choice.

In Fig.~\ref{fig:eugene} we show our zeroth-order, first-order and exact calculation
of this model.  We have verified that the exact calculation accurately 
reproduces the
results in Ref.~\cite{Chen:2006xjb}.

\bibliography{bifeature}

\begin{thebibliography}{27}%
\makeatletter
\providecommand \@ifxundefined [1]{%
 \@ifx{#1\undefined}
}%
\providecommand \@ifnum [1]{%
 \ifnum #1\expandafter \@firstoftwo
 \else \expandafter \@secondoftwo
 \fi
}%
\providecommand \@ifx [1]{%
 \ifx #1\expandafter \@firstoftwo
 \else \expandafter \@secondoftwo
 \fi
}%
\providecommand \natexlab [1]{#1}%
\providecommand \enquote  [1]{``#1''}%
\providecommand \bibnamefont  [1]{#1}%
\providecommand \bibfnamefont [1]{#1}%
\providecommand \citenamefont [1]{#1}%
\providecommand \href@noop [0]{\@secondoftwo}%
\providecommand \href [0]{\begingroup \@sanitize@url \@href}%
\providecommand \@href[1]{\@@startlink{#1}\@@href}%
\providecommand \@@href[1]{\endgroup#1\@@endlink}%
\providecommand \@sanitize@url [0]{\catcode `\\12\catcode `\$12\catcode
  `\&12\catcode `\#12\catcode `\^12\catcode `\_12\catcode `\%12\relax}%
\providecommand \@@startlink[1]{}%
\providecommand \@@endlink[0]{}%
\providecommand \url  [0]{\begingroup\@sanitize@url \@url }%
\providecommand \@url [1]{\endgroup\@href {#1}{\urlprefix }}%
\providecommand \urlprefix  [0]{URL }%
\providecommand \Eprint [0]{\href }%
\@ifxundefined \urlstyle {%
  \providecommand \doi  [0]{\begingroup \@sanitize@url \@doi}%
  \providecommand \@doi [1]{\endgroup \@@startlink {\doibase
  #1}doi:\discretionary {}{}{}#1\@@endlink }%
}{%
  \providecommand \doi  [0]{doi:\discretionary{}{}{}\begingroup
  \urlstyle{rm}\Url }%
}%
\providecommand \doibase [0]{http://dx.doi.org/}%
\providecommand \Doi [0]{\begingroup \@sanitize@url \@Doi }%
\providecommand \@Doi  [1]{\endgroup\@@startlink{\doibase#1}\@@Doi}%
\providecommand \@@Doi [1]{#1\@@endlink}%
\providecommand \selectlanguage [0]{\@gobble}%
\providecommand \bibinfo  [0]{\@secondoftwo}%
\providecommand \bibfield  [0]{\@secondoftwo}%
\providecommand \translation [1]{[#1]}%
\providecommand \BibitemOpen [0]{}%
\providecommand \bibitemStop [0]{}%
\providecommand \bibitemNoStop [0]{.\EOS\space}%
\providecommand \EOS [0]{\spacefactor3000\relax}%
\providecommand \BibitemShut  [1]{\csname bibitem#1\endcsname}%
\bibitem [{\citenamefont {Chen}\ \emph {et~al.}(2007)\citenamefont {Chen},
  \citenamefont {Easther},\ and\ \citenamefont {Lim}}]{Chen:2006xjb}%
  \BibitemOpen
  \bibfield  {author} {\bibinfo {author} {\bibfnamefont {X.}~\bibnamefont
  {Chen}}, \bibinfo {author} {\bibfnamefont {R.}~\bibnamefont {Easther}}, \
  and\ \bibinfo {author} {\bibfnamefont {E.~A.}\ \bibnamefont {Lim}},\ }\Doi
  {10.1088/1475-7516/2007/06/023} {\bibfield  {journal} {\bibinfo  {journal}
  {JCAP},\ }\textbf {\bibinfo {volume} {0706}},\ \bibinfo {pages} {023}
  (\bibinfo {year} {2007})},\ \Eprint {http://arxiv.org/abs/astro-ph/0611645}
  {arXiv:astro-ph/0611645 [astro-ph]} \BibitemShut {NoStop}%
\bibitem [{\citenamefont {Chen}\ \emph {et~al.}(2008)\citenamefont {Chen},
  \citenamefont {Easther},\ and\ \citenamefont {Lim}}]{Chen:2008wn}%
  \BibitemOpen
  \bibfield  {author} {\bibinfo {author} {\bibfnamefont {X.}~\bibnamefont
  {Chen}}, \bibinfo {author} {\bibfnamefont {R.}~\bibnamefont {Easther}}, \
  and\ \bibinfo {author} {\bibfnamefont {E.~A.}\ \bibnamefont {Lim}},\ }\Doi
  {10.1088/1475-7516/2008/04/010} {\bibfield  {journal} {\bibinfo  {journal}
  {JCAP},\ }\textbf {\bibinfo {volume} {0804}},\ \bibinfo {pages} {010}
  (\bibinfo {year} {2008})},\ \Eprint {http://arxiv.org/abs/0801.3295}
  {arXiv:0801.3295 [astro-ph]} \BibitemShut {NoStop}%
\bibitem [{\citenamefont {Stewart}(2002)}]{Stewart:2001cd}%
  \BibitemOpen
  \bibfield  {author} {\bibinfo {author} {\bibfnamefont {E.~D.}\ \bibnamefont
  {Stewart}},\ }\Doi {10.1103/PhysRevD.65.103508} {\bibfield  {journal}
  {\bibinfo  {journal} {Phys. Rev.},\ }\textbf {\bibinfo {volume} {D65}},\
  \bibinfo {pages} {103508} (\bibinfo {year} {2002})},\ \Eprint
  {http://arxiv.org/abs/astro-ph/0110322} {arXiv:astro-ph/0110322} \BibitemShut
  {NoStop}%
\bibitem [{\citenamefont {Dvorkin}\ and\ \citenamefont
  {Hu}(2010){\natexlab{a}}}]{Dvorkin:2009ne}%
  \BibitemOpen
  \bibfield  {author} {\bibinfo {author} {\bibfnamefont {C.}~\bibnamefont
  {Dvorkin}}\ and\ \bibinfo {author} {\bibfnamefont {W.}~\bibnamefont {Hu}},\
  }\Doi {10.1103/PhysRevD.81.023518} {\bibfield  {journal} {\bibinfo  {journal}
  {Phys. Rev.},\ }\textbf {\bibinfo {volume} {D81}},\ \bibinfo {pages} {023518}
  (\bibinfo {year} {2010}{\natexlab{a}})},\ \Eprint
  {http://arxiv.org/abs/0910.2237} {arXiv:0910.2237 [astro-ph.CO]} \BibitemShut
  {NoStop}%
\bibitem [{\citenamefont {Starobinsky}(1992)}]{Starobinsky:1992ts}%
  \BibitemOpen
  \bibfield  {author} {\bibinfo {author} {\bibfnamefont {A.~A.}\ \bibnamefont
  {Starobinsky}},\ }\href@noop {} {\bibfield  {journal} {\bibinfo  {journal}
  {JETP Lett.},\ }\textbf {\bibinfo {volume} {55}},\ \bibinfo {pages} {489}
  (\bibinfo {year} {1992})}\BibitemShut {NoStop}%
\bibitem [{\citenamefont {Adams}\ \emph {et~al.}(1997)\citenamefont {Adams},
  \citenamefont {Ross},\ and\ \citenamefont {Sarkar}}]{Adams:1997de}%
  \BibitemOpen
  \bibfield  {author} {\bibinfo {author} {\bibfnamefont {J.~A.}\ \bibnamefont
  {Adams}}, \bibinfo {author} {\bibfnamefont {G.~G.}\ \bibnamefont {Ross}}, \
  and\ \bibinfo {author} {\bibfnamefont {S.}~\bibnamefont {Sarkar}},\ }\Doi
  {10.1016/S0550-3213(97)00431-8} {\bibfield  {journal} {\bibinfo  {journal}
  {Nucl.Phys.},\ }\textbf {\bibinfo {volume} {B503}},\ \bibinfo {pages} {405}
  (\bibinfo {year} {1997})},\ \Eprint {http://arxiv.org/abs/hep-ph/9704286}
  {arXiv:hep-ph/9704286 [hep-ph]} \BibitemShut {NoStop}%
\bibitem [{\citenamefont {Adams}\ \emph {et~al.}(2001)\citenamefont {Adams},
  \citenamefont {Cresswell},\ and\ \citenamefont {Easther}}]{Adams:2001vc}%
  \BibitemOpen
  \bibfield  {author} {\bibinfo {author} {\bibfnamefont {J.~A.}\ \bibnamefont
  {Adams}}, \bibinfo {author} {\bibfnamefont {B.}~\bibnamefont {Cresswell}}, \
  and\ \bibinfo {author} {\bibfnamefont {R.}~\bibnamefont {Easther}},\ }\Doi
  {10.1103/PhysRevD.64.123514} {\bibfield  {journal} {\bibinfo  {journal}
  {Phys. Rev.},\ }\textbf {\bibinfo {volume} {D64}},\ \bibinfo {pages} {123514}
  (\bibinfo {year} {2001})},\ \Eprint {http://arxiv.org/abs/astro-ph/0102236}
  {arXiv:astro-ph/0102236} \BibitemShut {NoStop}%
\bibitem [{\citenamefont {Peiris}\ \emph {et~al.}(2003)\citenamefont {Peiris}
  \emph {et~al.}}]{Peiris:2003ff}%
  \BibitemOpen
  \bibfield  {author} {\bibinfo {author} {\bibfnamefont {H.~V.}\ \bibnamefont
  {Peiris}} \emph {et~al.} (\bibinfo {collaboration} {WMAP}),\ }\Doi
  {10.1086/377228} {\bibfield  {journal} {\bibinfo  {journal} {Astrophys. J.
  Suppl.},\ }\textbf {\bibinfo {volume} {148}},\ \bibinfo {pages} {213}
  (\bibinfo {year} {2003})},\ \Eprint {http://arxiv.org/abs/astro-ph/0302225}
  {arXiv:astro-ph/0302225} \BibitemShut {NoStop}%
\bibitem [{\citenamefont {Covi}\ \emph {et~al.}(2006)\citenamefont {Covi},
  \citenamefont {Hamann}, \citenamefont {Melchiorri}, \citenamefont {Slosar},\
  and\ \citenamefont {Sorbera}}]{Covi:2006ci}%
  \BibitemOpen
  \bibfield  {author} {\bibinfo {author} {\bibfnamefont {L.}~\bibnamefont
  {Covi}}, \bibinfo {author} {\bibfnamefont {J.}~\bibnamefont {Hamann}},
  \bibinfo {author} {\bibfnamefont {A.}~\bibnamefont {Melchiorri}}, \bibinfo
  {author} {\bibfnamefont {A.}~\bibnamefont {Slosar}}, \ and\ \bibinfo {author}
  {\bibfnamefont {I.}~\bibnamefont {Sorbera}},\ }\Doi
  {10.1103/PhysRevD.74.083509} {\bibfield  {journal} {\bibinfo  {journal}
  {Phys. Rev.},\ }\textbf {\bibinfo {volume} {D74}},\ \bibinfo {pages} {083509}
  (\bibinfo {year} {2006})},\ \Eprint {http://arxiv.org/abs/astro-ph/0606452}
  {arXiv:astro-ph/0606452} \BibitemShut {NoStop}%
\bibitem [{\citenamefont {Hamann}\ \emph {et~al.}(2007)\citenamefont {Hamann},
  \citenamefont {Covi}, \citenamefont {Melchiorri},\ and\ \citenamefont
  {Slosar}}]{Hamann:2007pa}%
  \BibitemOpen
  \bibfield  {author} {\bibinfo {author} {\bibfnamefont {J.}~\bibnamefont
  {Hamann}}, \bibinfo {author} {\bibfnamefont {L.}~\bibnamefont {Covi}},
  \bibinfo {author} {\bibfnamefont {A.}~\bibnamefont {Melchiorri}}, \ and\
  \bibinfo {author} {\bibfnamefont {A.}~\bibnamefont {Slosar}},\ }\Doi
  {10.1103/PhysRevD.76.023503} {\bibfield  {journal} {\bibinfo  {journal}
  {Phys. Rev.},\ }\textbf {\bibinfo {volume} {D76}},\ \bibinfo {pages} {023503}
  (\bibinfo {year} {2007})},\ \Eprint {http://arxiv.org/abs/astro-ph/0701380}
  {arXiv:astro-ph/0701380} \BibitemShut {NoStop}%
\bibitem [{\citenamefont {Mortonson}\ \emph {et~al.}(2009)\citenamefont
  {Mortonson}, \citenamefont {Dvorkin}, \citenamefont {Peiris},\ and\
  \citenamefont {Hu}}]{Mortonson:2009qv}%
  \BibitemOpen
  \bibfield  {author} {\bibinfo {author} {\bibfnamefont {M.~J.}\ \bibnamefont
  {Mortonson}}, \bibinfo {author} {\bibfnamefont {C.}~\bibnamefont {Dvorkin}},
  \bibinfo {author} {\bibfnamefont {H.~V.}\ \bibnamefont {Peiris}}, \ and\
  \bibinfo {author} {\bibfnamefont {W.}~\bibnamefont {Hu}},\ }\Doi
  {10.1103/PhysRevD.79.103519} {\bibfield  {journal} {\bibinfo  {journal}
  {Phys. Rev.},\ }\textbf {\bibinfo {volume} {D79}},\ \bibinfo {pages} {103519}
  (\bibinfo {year} {2009})},\ \Eprint {http://arxiv.org/abs/0903.4920}
  {arXiv:0903.4920 [astro-ph.CO]} \BibitemShut {NoStop}%
\bibitem [{\citenamefont {Hailu}\ and\ \citenamefont
  {Tye}(2007)}]{Hailu:2006uj}%
  \BibitemOpen
  \bibfield  {author} {\bibinfo {author} {\bibfnamefont {G.}~\bibnamefont
  {Hailu}}\ and\ \bibinfo {author} {\bibfnamefont {S.-H.}\ \bibnamefont
  {Tye}},\ }\Doi {10.1088/1126-6708/2007/08/009} {\bibfield  {journal}
  {\bibinfo  {journal} {JHEP},\ }\textbf {\bibinfo {volume} {0708}},\ \bibinfo
  {pages} {009} (\bibinfo {year} {2007})},\ \Eprint
  {http://arxiv.org/abs/hep-th/0611353} {arXiv:hep-th/0611353 [hep-th]}
  \BibitemShut {NoStop}%
\bibitem [{\citenamefont {Bean}\ \emph {et~al.}(2008)\citenamefont {Bean},
  \citenamefont {Chen}, \citenamefont {Hailu}, \citenamefont {Tye},\ and\
  \citenamefont {Xu}}]{Bean:2008na}%
  \BibitemOpen
  \bibfield  {author} {\bibinfo {author} {\bibfnamefont {R.}~\bibnamefont
  {Bean}}, \bibinfo {author} {\bibfnamefont {X.}~\bibnamefont {Chen}}, \bibinfo
  {author} {\bibfnamefont {G.}~\bibnamefont {Hailu}}, \bibinfo {author}
  {\bibfnamefont {S.-H.}\ \bibnamefont {Tye}}, \ and\ \bibinfo {author}
  {\bibfnamefont {J.}~\bibnamefont {Xu}},\ }\Doi
  {10.1088/1475-7516/2008/03/026} {\bibfield  {journal} {\bibinfo  {journal}
  {JCAP},\ }\textbf {\bibinfo {volume} {0803}},\ \bibinfo {pages} {026}
  (\bibinfo {year} {2008})},\ \Eprint {http://arxiv.org/abs/0802.0491}
  {arXiv:0802.0491 [hep-th]} \BibitemShut {NoStop}%
\bibitem [{\citenamefont {Abolhasani}\ \emph {et~al.}(2010)\citenamefont
  {Abolhasani}, \citenamefont {Firouzjahi},\ and\ \citenamefont
  {Namjoo}}]{Abolhasani:2010kn}%
  \BibitemOpen
  \bibfield  {author} {\bibinfo {author} {\bibfnamefont {A.~A.}\ \bibnamefont
  {Abolhasani}}, \bibinfo {author} {\bibfnamefont {H.}~\bibnamefont
  {Firouzjahi}}, \ and\ \bibinfo {author} {\bibfnamefont {M.~H.}\ \bibnamefont
  {Namjoo}},\ }\href@noop {} { (\bibinfo {year} {2010})},\ \Eprint
  {http://arxiv.org/abs/1010.6292} {arXiv:1010.6292 [astro-ph.CO]} \BibitemShut
  {NoStop}%
\bibitem [{\citenamefont {Achucarro}\ \emph {et~al.}(2011)\citenamefont
  {Achucarro}, \citenamefont {Gong}, \citenamefont {Hardeman}, \citenamefont
  {Palma},\ and\ \citenamefont {Patil}}]{Achucarro:2010da}%
  \BibitemOpen
  \bibfield  {author} {\bibinfo {author} {\bibfnamefont {A.}~\bibnamefont
  {Achucarro}}, \bibinfo {author} {\bibfnamefont {J.-O.}\ \bibnamefont {Gong}},
  \bibinfo {author} {\bibfnamefont {S.}~\bibnamefont {Hardeman}}, \bibinfo
  {author} {\bibfnamefont {G.~A.}\ \bibnamefont {Palma}}, \ and\ \bibinfo
  {author} {\bibfnamefont {S.~P.}\ \bibnamefont {Patil}},\ }\Doi
  {10.1088/1475-7516/2011/01/030} {\bibfield  {journal} {\bibinfo  {journal}
  {JCAP},\ }\textbf {\bibinfo {volume} {1101}},\ \bibinfo {pages} {030}
  (\bibinfo {year} {2011})},\ \Eprint {http://arxiv.org/abs/1010.3693}
  {arXiv:1010.3693 [hep-ph]} \BibitemShut {NoStop}%
\bibitem [{\citenamefont {Joy}\ \emph {et~al.}(2008)\citenamefont {Joy},
  \citenamefont {Sahni},\ and\ \citenamefont {Starobinsky}}]{Joy:2007na}%
  \BibitemOpen
  \bibfield  {author} {\bibinfo {author} {\bibfnamefont {M.}~\bibnamefont
  {Joy}}, \bibinfo {author} {\bibfnamefont {V.}~\bibnamefont {Sahni}}, \ and\
  \bibinfo {author} {\bibfnamefont {A.~A.}\ \bibnamefont {Starobinsky}},\ }\Doi
  {10.1103/PhysRevD.77.023514} {\bibfield  {journal} {\bibinfo  {journal}
  {Phys.Rev.},\ }\textbf {\bibinfo {volume} {D77}},\ \bibinfo {pages} {023514}
  (\bibinfo {year} {2008})},\ \Eprint {http://arxiv.org/abs/0711.1585}
  {arXiv:0711.1585 [astro-ph]} \BibitemShut {NoStop}%
\bibitem [{\citenamefont {Hotchkiss}\ and\ \citenamefont
  {Sarkar}(2010)}]{Hotchkiss:2009pj}%
  \BibitemOpen
  \bibfield  {author} {\bibinfo {author} {\bibfnamefont {S.}~\bibnamefont
  {Hotchkiss}}\ and\ \bibinfo {author} {\bibfnamefont {S.}~\bibnamefont
  {Sarkar}},\ }\Doi {10.1088/1475-7516/2010/05/024} {\bibfield  {journal}
  {\bibinfo  {journal} {JCAP},\ }\textbf {\bibinfo {volume} {1005}},\ \bibinfo
  {pages} {024} (\bibinfo {year} {2010})},\ \Eprint
  {http://arxiv.org/abs/0910.3373} {arXiv:0910.3373 [astro-ph.CO]} \BibitemShut
  {NoStop}%
\bibitem [{\citenamefont {Nakashima}\ \emph {et~al.}(2010)\citenamefont
  {Nakashima}, \citenamefont {Saito}, \citenamefont {Takamizu},\ and\
  \citenamefont {Yokoyama}}]{Nakashima:2010sa}%
  \BibitemOpen
  \bibfield  {author} {\bibinfo {author} {\bibfnamefont {M.}~\bibnamefont
  {Nakashima}}, \bibinfo {author} {\bibfnamefont {R.}~\bibnamefont {Saito}},
  \bibinfo {author} {\bibfnamefont {Y.-i.}\ \bibnamefont {Takamizu}}, \ and\
  \bibinfo {author} {\bibfnamefont {J.}~\bibnamefont {Yokoyama}},\ }\href@noop
  {} { (\bibinfo {year} {2010})},\ \Eprint {http://arxiv.org/abs/1009.4394}
  {arXiv:1009.4394 [astro-ph.CO]} \BibitemShut {NoStop}%
\bibitem [{\citenamefont {Maldacena}(2003)}]{Maldacena:2002vr}%
  \BibitemOpen
  \bibfield  {author} {\bibinfo {author} {\bibfnamefont {J.~M.}\ \bibnamefont
  {Maldacena}},\ }\href@noop {} {\bibfield  {journal} {\bibinfo  {journal}
  {JHEP},\ }\textbf {\bibinfo {volume} {0305}},\ \bibinfo {pages} {013}
  (\bibinfo {year} {2003})},\ \Eprint {http://arxiv.org/abs/astro-ph/0210603}
  {arXiv:astro-ph/0210603 [astro-ph]} \BibitemShut {NoStop}%
\bibitem [{\citenamefont {Weinberg}(2005)}]{Weinberg:2005vy}%
  \BibitemOpen
  \bibfield  {author} {\bibinfo {author} {\bibfnamefont {S.}~\bibnamefont
  {Weinberg}},\ }\Doi {10.1103/PhysRevD.72.043514} {\bibfield  {journal}
  {\bibinfo  {journal} {Phys.Rev.},\ }\textbf {\bibinfo {volume} {D72}},\
  \bibinfo {pages} {043514} (\bibinfo {year} {2005})},\ \Eprint
  {http://arxiv.org/abs/hep-th/0506236} {arXiv:hep-th/0506236 [hep-th]}
  \BibitemShut {NoStop}%
\bibitem [{\citenamefont {Adshead}\ \emph {et~al.}()\citenamefont {Adshead},
  \citenamefont {Burrage}, \citenamefont {Ribeiro},\ and\ \citenamefont
  {Seery}}]{AdsheadSeery}%
  \BibitemOpen
  \bibfield  {author} {\bibinfo {author} {\bibfnamefont {P.}~\bibnamefont
  {Adshead}}, \bibinfo {author} {\bibfnamefont {C.}~\bibnamefont {Burrage}},
  \bibinfo {author} {\bibfnamefont {R.~H.}\ \bibnamefont {Ribeiro}}, \ and\
  \bibinfo {author} {\bibfnamefont {D.}~\bibnamefont {Seery}},\ }\href@noop {}
  {}\bibinfo {note} {{in prep.}}\BibitemShut {Stop}%
\bibitem [{\citenamefont {Creminelli}\ and\ \citenamefont
  {Zaldarriaga}(2004)}]{Creminelli:2004yq}%
  \BibitemOpen
  \bibfield  {author} {\bibinfo {author} {\bibfnamefont {P.}~\bibnamefont
  {Creminelli}}\ and\ \bibinfo {author} {\bibfnamefont {M.}~\bibnamefont
  {Zaldarriaga}},\ }\Doi {10.1088/1475-7516/2004/10/006} {\bibfield  {journal}
  {\bibinfo  {journal} {JCAP},\ }\textbf {\bibinfo {volume} {0410}},\ \bibinfo
  {pages} {006} (\bibinfo {year} {2004})},\ \Eprint
  {http://arxiv.org/abs/astro-ph/0407059} {arXiv:astro-ph/0407059 [astro-ph]}
  \BibitemShut {NoStop}%
\bibitem [{\citenamefont {Hu}(2000)}]{Hu:2000ee}%
  \BibitemOpen
  \bibfield  {author} {\bibinfo {author} {\bibfnamefont {W.}~\bibnamefont
  {Hu}},\ }\Doi {10.1103/PhysRevD.62.043007} {\bibfield  {journal} {\bibinfo
  {journal} {Phys.Rev.},\ }\textbf {\bibinfo {volume} {D62}},\ \bibinfo {pages}
  {043007} (\bibinfo {year} {2000})},\ \Eprint
  {http://arxiv.org/abs/astro-ph/0001303} {arXiv:astro-ph/0001303 [astro-ph]}
  \BibitemShut {NoStop}%
\bibitem [{\citenamefont {Flauger}\ and\ \citenamefont
  {Pajer}(2011)}]{Flauger:2010ja}%
  \BibitemOpen
  \bibfield  {author} {\bibinfo {author} {\bibfnamefont {R.}~\bibnamefont
  {Flauger}}\ and\ \bibinfo {author} {\bibfnamefont {E.}~\bibnamefont
  {Pajer}},\ }\Doi {10.1088/1475-7516/2011/01/017} {\bibfield  {journal}
  {\bibinfo  {journal} {JCAP},\ }\textbf {\bibinfo {volume} {1101}},\ \bibinfo
  {pages} {017} (\bibinfo {year} {2011})},\ \Eprint
  {http://arxiv.org/abs/1002.0833} {arXiv:1002.0833 [hep-th]} \BibitemShut
  {NoStop}%
\bibitem [{\citenamefont {Leblond}\ and\ \citenamefont
  {Pajer}(2011)}]{Leblond:2010yq}%
  \BibitemOpen
  \bibfield  {author} {\bibinfo {author} {\bibfnamefont {L.}~\bibnamefont
  {Leblond}}\ and\ \bibinfo {author} {\bibfnamefont {E.}~\bibnamefont
  {Pajer}},\ }\Doi {10.1088/1475-7516/2011/01/035} {\bibfield  {journal}
  {\bibinfo  {journal} {JCAP},\ }\textbf {\bibinfo {volume} {1101}},\ \bibinfo
  {pages} {035} (\bibinfo {year} {2011})},\ \Eprint
  {http://arxiv.org/abs/1010.4565} {arXiv:1010.4565 [hep-th]} \BibitemShut
  {NoStop}%
\bibitem [{\citenamefont {Choe}\ \emph {et~al.}(2004)\citenamefont {Choe},
  \citenamefont {Gong},\ and\ \citenamefont {Stewart}}]{Choe:2004zg}%
  \BibitemOpen
  \bibfield  {author} {\bibinfo {author} {\bibfnamefont {J.}~\bibnamefont
  {Choe}}, \bibinfo {author} {\bibfnamefont {J.-O.}\ \bibnamefont {Gong}}, \
  and\ \bibinfo {author} {\bibfnamefont {E.~D.}\ \bibnamefont {Stewart}},\
  }\Doi {10.1088/1475-7516/2004/07/012} {\bibfield  {journal} {\bibinfo
  {journal} {JCAP},\ }\textbf {\bibinfo {volume} {0407}},\ \bibinfo {pages}
  {012} (\bibinfo {year} {2004})},\ \Eprint
  {http://arxiv.org/abs/hep-ph/0405155} {arXiv:hep-ph/0405155} \BibitemShut
  {NoStop}%
\bibitem [{\citenamefont {Dvorkin}\ and\ \citenamefont
  {Hu}(2010){\natexlab{b}}}]{DvoHu10a}%
  \BibitemOpen
  \bibfield  {author} {\bibinfo {author} {\bibfnamefont {C.}~\bibnamefont
  {Dvorkin}}\ and\ \bibinfo {author} {\bibfnamefont {W.}~\bibnamefont {Hu}},\
  }\Doi {10.1103/PhysRevD.82.043513} {\bibfield  {journal} {\bibinfo  {journal}
  {Phys. Rev.},\ }\textbf {\bibinfo {volume} {D82}},\ \bibinfo {pages} {043513}
  (\bibinfo {year} {2010}{\natexlab{b}})},\ \Eprint
  {http://arxiv.org/abs/1007.0215} {arXiv:1007.0215 [astro-ph.CO]} \BibitemShut
  {NoStop}%
\end{thebibliography}%

\end{document}